\documentclass[journal,final,twocolumn,10pt]{IEEEtran}
\usepackage{graphicx}
\usepackage{mhchem}
\usepackage{subfig}
\usepackage{url}
\usepackage{epstopdf}
\usepackage{amsmath,cuted,widetext,lipsum}
\usepackage{amssymb}
\usepackage{balance}
\usepackage{cite}
\usepackage{citesort}
\usepackage[usenames, dvipsnames]{color}
\usepackage{amsthm}
\begin{document}
\bstctlcite{IEEEexample:BSTcontrol}

\title{Analyzing the Impact of Molecular Re-Radiation on the MIMO Capacity in High-Frequency Bands}

\author{ \IEEEauthorblockN{Sayed Amir Hoseini, Ming Ding,~\IEEEmembership{Senior Member,~IEEE}, Mahbub Hassan,~\IEEEmembership{Senior Member,~IEEE} and Youjia Chen,~\IEEEmembership{Member,~IEEE}} 
		\thanks{Copyright (c) 2020 IEEE. Personal use of this material is permitted. However, permission to use this material for any other purposes must be obtained from the IEEE by sending a request to pubs-permissions@ieee.org.}
		\thanks{Sayed Amir Hoseini (corresponding author) is with the School of Engineering and Information Technology (SEIT), University of New South Wales (UNSW) at Canberra, Northcott Drive, Campbell, ACT 2600, Australia. e-mail: s.a.hoseini@unsw.edu.au.}
		\thanks{Ming Ding is with Data61, CSIRO, Eveleigh, NSW 2015, Australia. e-mail: Ming.Ding@data61.csiro.au.}
		\thanks{Mahbub Hassan is with the School of Computer Science and Engineering, University of New South Wales (UNSW), Sydney, Australia. E-mail: mahbub.hassan@unsw.edu.au.}
		\thanks{Youjia chen is with the College of Physics and Information Engineering, Fuzhou University, China. e-mail: youjia.chen@fzu.edu.cn.}
		\thanks{This work is supported by the Commonwealth Scientific and Industrial Research Organization (CSIRO) of Australia and the National Natural Science Foundation of China (NSFC Grant No. 61801119).}
}
\maketitle

\begin{abstract}
In this paper, we show how the absorption and re-radiation energy from molecules in the air can influence the Multiple Input Multiple Output (MIMO) performance in high-frequency bands, e.g., millimeter wave (mmWave) and terahertz. In more detail, some common atmosphere molecules, such as oxygen and water, can absorb and re-radiate energy in their natural resonance frequencies, such as 60\,GHz, 180\,GHz and 320\,GHz. Hence, when hit by electromagnetic waves, molecules will get excited and absorb energy, which leads to an extra path loss and is known as molecular attenuation. Meanwhile, the absorbed energy will be re-radiated towards a random direction with a random phase. These re-radiated waves also interfere with the signal transmission. Although, the molecular re-radiation was mostly considered as noise in literature, recent works show that it is correlated to the main signal and can be viewed as a composition of multiple delayed or scattered signals. Such a phenomenon can provide non-line-of-sight (NLoS) paths in an environment that lacks scatterers, which increases spatial multiplexing and thus greatly enhances the performance of MIMO systems. Therefore in this paper, we explore the scattering model and noise models of molecular re-radiation to characterize the channel transfer function of the NLoS channels created by atmosphere molecules. Our simulation results show that the re-radiation can increase MIMO capacity up to 3 folds in mmWave and 6 folds in terahertz for a set of realistic transmit power, distance, and antenna numbers. We also show that in the high SNR, the re-radiation makes the open-loop precoding viable, which is an alternative to beamforming to avoid beam alignment sensitivity in high mobility applications.

\end{abstract}

\section{ Introduction}\label{sec:introduction}

The relentless growth of data traffic is presenting significant challenges for wireless network providers. 
Fixed Wireless Access (FWA) connections are expected to increase three folds by the end of 2025 
while the total mobile data volume will grow more than fourfold during the same period~\cite{Ericsson2020}, requiring enormous capacity enhancements in wireless communication systems. 5G systems aim to provide a peak data rate of 10 Gbps per user \cite{karjalainen2014challenges} while 6G is expected to enhance capacity 10–100 times over 5G \cite{ZHOU2020253}. In addition to cellular network capacity demands, local wireless networks are expected to support Tbps data rates \cite{saad2018beyond} to realize super bandwidth-hungry applications such as wireless virtual reality (VR)~\cite{du2020mec}. 

The major portion of the wireless communication traffic currently uses frequencies below 6\,GHz. Unfortunately, this part of the spectrum is currently highly saturated and will soon become overloaded. Thus, in spite of the very efficient spectrum use in recent wireless standards, it is becoming necessary to utilize the higher frequency bands above 6\,GHz to realistically accommodate the future traffic growth. In particular, largely unused parts of the spectrum in the range of 30-300\,GHz, a.k.a. mmWave band, is currently being considered for use in 5G/6G systems~\cite{Zhou2019-6G}. To achieve extremely high bit rates, the terahertz band in the range of 0.1-10\,THz is also being considered by many research groups~\cite{akyildiz2018combating} for short-range applications such as nanoscale sensor networks~\cite{zarepour2017semon}, wireless on-chip communications, and wireless personal area networks~\cite{Akyildiz201416}.

Unfortunately, such high-frequency bands not only suffer from high path loss but also encounter severe frequency-selective molecular absorption, which is not observable in sub-6\,GHz frequencies. Recent studies have confirmed that the molecules in the communication medium can absorb
significant amounts of wireless signal energy if excited in their natural resonance frequencies~\cite{Kokkoniemi2015discuss,Jornet2014a,Akyildiz201416,carbondi2012}.
For a normal atmosphere, oxygen and water molecules are the major players in molecular absorption with their natural resonance frequencies at around 60\,GHz, 180\,GHz, 320\,GHz, and so on. Interestingly, according to fundamental physics, the atmospheric molecules not only absorb the energy from the high-frequency electromagnetic waves but they also re-radiate some of the absorbed energies at the same frequencies shortly after their absorption.  
Although this re-radiation is basically a source of additional 
noise~\cite{Noise_1986,Jornet2014a} for high-frequency bands, detailed theoretical and experimental studies have revealed that it is highly correlated to the main signal~\cite{jornet2013fundamentals,harde1991coherent} and hence can be modeled as a scattered copy of the original signal~\cite{Kokkoniemi2015discuss}. 

The absorption and re-radiation of the electromagnetic wave are not limited to oxygen and water. An experiment has shown that nitrogen dioxide molecules absorb energy from a terahertz pulse and re-radiate a distorted copy of the pulse shortly after~\cite{harde1991coherent}. Molecules of carbon dioxide absorb photon energy in infrared frequency and shortly re-radiate them in the same band~\cite{carbondi2012}. In the ultraviolet spectrum, ozone is the primary constituent that absorbs and re-radiates electromagnetic waves~\cite{mori2016ultraviolet}. The positive role of the molecular re-radiation, known as molecular scattering, in ultraviolet communication is widely studied theoretically and experimentally where a practical non-line-of-sight (NLoS) channel is created by molecules and aerosols~\cite{mori2016ultraviolet,ding2010characterization,xu2007experimental}. 

Therefore, as a preliminary investigation into taking advantage of potential NLoS signal reception due to molecular re-radiation, we conduct a theoretical study on the performance impacts of the re-radiation behavior of oxygen and water vapor molecules on MIMO systems. In this paper, the molecular re-radiation is analyzed under two different approaches: (1) re-radiation is considered as random noise, or (2) re-radiation is treated as correlated scattering. As widely known, noise is harmful for both single antenna and multi-antenna channels, while scattering is viewed as a very important factor to provide spatial multiplexing in MIMO. Since the molecular absorption intensity depends on the resonance frequencies of the molecules in the atmosphere,
the energy absorption and re-radiation are not flat in the mmWave/terahertz spectrum. Thus, the wireless channel experiences a frequency-selective noise or scattering along with the frequency-selective attenuation \cite{hoseini2016diurnal}.

Based on the aforementioned interpretation of molecular re-radiation, we propose the conjecture that the MIMO performance in the mmWave and terahertz bands would also be frequency-selective.
We verify our conjecture via theoretical studies and computer simulations, which show that the MIMO capacity increases dramatically in specific high-absorption frequency windows thanks to the ubiquitous existence of oxygen and water molecules. The intuition of our theoretical study is that the channel molecules can radiate a distorted copy of the signal with an additional random phase, which equivalently creates a richer scattering environment and consequently the potential to improve the line-of-sight (LoS) MIMO capacity.

Our discovery can fundamentally change our understanding of the relationship between the MIMO capacity and the frequency spectrum. While the literature has extensively studied frequency selectivity of mmWave/terahertz communications \cite{yuan2018hybrid,gonzalez2018channel,gonzalez2016frequency,yuan2018low}, our study provides a new perspective of frequency selectivity in the context of MIMO performance in these emerging spectrum bands.
Importantly, our results reveal that several mmWave and terahertz frequency windows can serve as valuable spectrum for high-efficiency MIMO communications, which may in turn shift the paradigm of research, standardization, and implementation in the field of high-frequency communications. In the future, it may be even possible to engineer mechanics at the transmitter/receiver side to change the vapor and/or oxygen densities in the medium, which would in turn proactively change the mmWave/terahertz channel environment to boost system capacity. 

We have published parts of the results in our previous papers~\cite{hoseini2017mmWave,hoseini2017massive,hoseini2017thesis}. In~\cite{hoseini2017mmWave}, we simulated a simple MIMO system in the mmWave band and showed the impact of re-radiation on the performance of the multiplexing technique. In the second step~\cite{hoseini2017massive}, we compared beamforming and multiplexing techniques in a massive MIMO system in the terahertz band. The main contributions of this paper can be summarized as follows:
\begin{itemize}
	\item We analyze the theoretical upper bound and lower bound of MIMO capacity under the scattering model of molecular re-radiation.
	\item For the scattering model of re-radiation, we re-define the Rician K-factor with the absorption coefficient. 
	\item We investigate the extended multiplexing technique with the optimal precoding matrix and power allocation scheme which owns beamforming and multiplexing benefits together. We show it can outperform the beamforming technique in the presence of molecular re-radiation. 

\end{itemize}

The rest of the paper is structured as follows.
In Section~\ref{sec:Related}, 
a brief literature review is presented.
In Section~\ref{sec:model}, 
we firstly present the molecular absorption model,
and then the fundamental theory of MIMO capacity is reviewed.
Section~\ref{sec:analysis} analyzes MIMO capacity versus the molecular re-radiation, 
followed by simulation results and insightful discussions in Section~\ref{sec:simulSetup}.
The conclusion is presented in Section~\ref{sec:con} and the future research directions are discussed in Section~\ref{sec:fut}.


\section{ Related Work}\label{sec:Related}

Molecular absorption and re-radiation are well-known phenomena,
which have been studied extensively in the literature for mmWave and Terahertz bands~\cite{Noise_1986,jornet2013fundamentals,Akdeniz2014Capacity,Akyildiz2010c,ICC2015_Noise}. 
The majority of these works have considered the re-radiated energy as a form of random noise. The molecular absorption noise has been studied in the literature as early as 1986 when the authors in \cite{Noise_1986} proposed a model for the sky atmospheric noise for frequencies higher than 18GHz. 
However, they mostly investigated the sky noise where the magnitude is not influenced by the amplitude of the transmitted signal. This type of molecular noise has been referred to as brightness noise in ITU recommendation \cite{noise2001itu} where it is modeled as antenna noise. Moreover, in \cite{Noise_1986,jornet2013fundamentals,Pierobon2013_Routing}, it has been assumed that the intensity of the transmitted power affects the molecular noise, which is known as self-induced noise. Authors in~\cite{jornet2013fundamentals} have proposed that the self-induced noise should be correlated to the original signal,
and modeled it as scattering from multiple virtual NLoS paths~\cite{Kokkoniemi2015discuss}. However, the proposed model in~\cite{Kokkoniemi2015discuss} only characterized the power delay profile of the channel and did not include the phase information of re-radiated waves. Furthermore, authors in \cite{Kokkoniemi2015discuss} have presented a comprehensive discussion on alternative interpretations of molecular re-radiation as noise or scattering.

An interesting experimental study has shown that the medium molecules can absorb and re-radiate a subpicosecond terahertz pulse \cite{harde1991coherent}. More specifically, the authors have excited \ce{N_2O} molecules in a relatively short channel with a terahertz pulse and observed that a train of subpicosecond pulses was re-radiated by channel molecules. The relative amplitude of the first re-radiated pulse was measured as one tenth 
of the excitation pulse.

Several studies have investigated the effect of atmospheric conditions on the mmWave channel \cite{Siles2015Atmospheric,Frey1999Atmosphere,Liebe1985moise,Yong2015Rainfall,hoseini2016diurnal}. In particular, it was found that heavy rainfall would severely attenuate mmWave communications~\cite{Zhouyue2011mmwave}. The rainfall was modeled and evaluated in \cite{Yong2015Rainfall} where the authors found that mild rainfall could in fact have a positive effect on MIMO performance as the raindrops were found to act as scatterers and improve the MIMO multiplexing gain. Interestingly, they also found that with increasing rain, the attenuation dominates the scattering effects of the raindrops, which results in capacity degradation under heavy rain.

In \cite{mori2016ultraviolet,ding2010characterization,xu2007experimental}, authors have discussed the possibility of NLoS communication in the Ultra-Violate (UV) spectrum thanks to tropospheric molecular and aerosols scattering. In the UV spectrum, ozone is the primary constituent that absorbs and scatters the electromagnetic wave. All these works have conducted numerical and experimental analysis to show that the NLoS scattering by air can be an alternative to LoS communication where UV LoS channel is vulnerable to blockage and shadowing. However, to the best of our knowledge,
there is no previous work on studying the impact of the molecular absorption and re-radiation on the MIMO capacity.

\section{Channel Model and MIMO Capacity} \label{sec:model}

In this section, we present the channel model that considers the effect of molecular absorption and re-radiation on the electromagnetic waves propagating through the channel. Later, MIMO capacity is analyzed for such channels.

\subsection{Molecular Absorption Coefficient}
The absorption of a given molecule 
is characterized by absorption coefficient $k_i(f)$ at frequency $f$, which varies with pressure and temperature of the environment.
The molecular absorption coefficients of many chemical species for different pressure and temperature are available from the publicly accessible databases such as \emph{HITRAN}~\cite{Rothman2012Database} and \textit{NIST Atomic Spectra}~\cite{NIST}. To model molecular absorption of a communication channel,
let us assume that the wireless channel is a medium consisting of $N$ chemical species where $m_1, m_2, ..., m_N$ are the mole fraction per volume of spices.
The \emph{medium absorption coefficient},
i.e., $k(f)$,
at frequency $f$ is, therefore, a weighted sum of the molecular absorption coefficients in the medium~\cite{Jornet2014a}, which can be formulated as
\begin{equation}
k(f) = \sum_{i=1}^{N} m_i k_{i}(f),
\label{eq:Kf}
\end{equation}
where $k_{i}(f)$ is the molecular absorption coefficient of species $i$. It should be noted that climate conditions, such as summer versus winter, or even the weather changes during the day would affect the absorption coefficients in the mmWave/terahertz band~\cite{hoseini2016diurnal}. In this work, we extract $k_{i}(f)$ from HITRAN~\cite{Rothman2012Database} for some predefined standard atmospheric conditions as shown
in Table~\ref{tab:gasMix}.

\subsection{Attenuation of Radio Signal}

The attenuation of the electromagnetic signal at the mmWave/terahertz frequencies is due to free space path loss (FSPL) and molecular absorption~\cite{mmWavepathloss}.
The total attenuation at frequency $f$ at a distance of $d$ from the radio transmitter can be written as
\begin{equation}
A(f,d) = A_{\rm FSPL}(f, d) \times A_{\rm a}(f,d),
\label{eq:Atten_Total}
\end{equation}
where $A_{\rm FSPL}(f,d)$ and $A_{\rm a}(f,d)$ denote the attenuation due to FSPL and molecular absorption, respectively. In more detail,
the FSPL attenuation is given by
\begin{equation}
 A_{\rm FSPL}(f, d)  = \left( \frac{4 \pi f d}{c}\right) ^2,
 \label{eq:Atten_Spread}
\end{equation}
where $c$ is the speed of light. The attenuation due to molecular absorption is characterized as~\cite{mmWavepathloss}
\begin{equation}
A_{\rm a}(f,d) = e^{k(f) \times d},
\label{eq:Atten_abs}
\end{equation}
where $k(f)$ is the absorption coefficient of the medium at frequency $f$. Thus,
the received power for LoS path is obtained as
\begin{eqnarray}
  \nonumber P_{{\rm r,LoS}}(f,d) &=& \frac{P_{t}(f)}{A(f,d)} \\
    &=& P_{t}(f)\times\left(\frac{c}{4\pi fd}\right)^{2}\times e^{-k(f)\times d}.
    \label{eq:Atten_pr1}
\end{eqnarray}

\subsection{Molecular Re-radiation}
\label{sec:molecular-reradiation}
The absorbed energy excites the channel molecules by increasing their vibrational-rotational energy levels~\cite{carbondi2012}.
The excitement is temporary and the vibrational-rotational energy returns to a steady state where the absorbed energy is re-radiated in the same frequency. 
In the literature, the re-radiated energy is commonly assumed as an additional source of noise, often referred to as the molecular noise~\cite{Akyildiz201416}.
As different molecule species exhibit different resonance frequencies, the power spectral density of the molecular noise, $N_{\rm a}$, is not flat over its spectrum. Generally speaking, both the atmospheric noise, $N_{atm}$, and the self-induced noise, $N_{si}$, contribute to the molecular noise as investigated in~\cite{Noise_1986,Jornet2014a}:
\begin{align}
&N_{\rm a}(f,d)= N_{atm}(f,d)+N_{si}(f,d), \label{eq:NoisePDS}\\
&N_{atm}(f,d)=lim_{d\to\infty} (k_B T_0 (1-e^{-k(f) d})) \Big( \frac{c}{\sqrt{4\pi} f} \Big)^2, \label{eq:Noise01}\\
&N_{si}(f,d)=P_t(f)(1-e^{-k(f) d}) \Big(\frac{c}{{4\pi f d}}\Big)^2, \label{eq:Noise02}
\end{align}
where
$T_0$ is the reference temperature ($ 296K) $,
$k_B$ is the Boltzmann constant,
$P_t(f)$ is the power spectral density of the transmitted signal and $c$ is the speed of light.
The first term in~\eqref{eq:NoisePDS},
which is also called sky noise~\eqref{eq:Noise01}, is independent of the signal wave.
However,
the self-induced noise~\eqref{eq:Noise02} is highly correlated with the signal wave~\cite{jornet2013fundamentals},
and can be considered as a distorted copy of the signal wave.
Thus, the received power of the signal re-radiated by the molecules can be expressed as follows:
\begin{equation}
 P_{\rm r,a}(f,d)=P_t(f)(1-e^{-k(f) d}) \Big(\frac{c}{{4\pi f d}}\Big)^2.
\label{eq:Atten_pr2}
\end{equation}

Since the phase of the re-radiated wave depends on the phase of molecular vibration,
which varies from molecules to molecules~\cite{barron2004molecular},
the received power is actually affected by a large number of phase-independent re-radiated photons.
Thus, the phase for the received signal, $\beta$, is assumed to be uniformly distributed in the range $[0, 2\pi)$, with its power given by~\eqref{eq:Atten_pr2}.

\subsection{Channel Transfer Function}

\label{sec:Hfunction}
The channel transfer function of an LoS channel is given by
\begin{equation}
\begin{aligned}
{h}_{\rm LoS}(f,d) &=\sqrt{ \left( \frac{c}{4 \pi f d}\right) ^2 e^{-k(f) \times d}} \times e^ {j2\pi\frac{d}{\lambda}} \\
			   &= \left( \frac{c}{4 \pi f d}\right) e^{-k(f) \times \frac{d}{2}} \times e^ {j2\pi\frac{d}{\lambda}}.
\label{eq:Hfunc_1}
\end{aligned}
\end{equation}

Then,
the partial channel transfer function resulting from the molecular absorption and excluding the LoS component can be represented by
\begin{equation}
\begin{aligned}
{h}_{\rm a}(f,d)&=\sqrt{(1-e^{-k(f) d}) \Big(\frac{c}{{4\pi f d}}\Big)^2}\times e^ {j2\pi\beta}\\
 			 &=(1-e^{-k(f) d})^{\frac{1}{2}} \Big(\frac{c}{{4\pi f d}}\Big) \times e^ {j2\pi\beta}.
\label{eq:Hfunc_2}
\end{aligned}
\end{equation}

Hence,
the total channel transfer function is the superposition of the partial channel transfer functions,
which is written as
\begin{eqnarray}
 \nonumber {h}(f,d) \hspace{-0.2cm}&=&\hspace{-0.2cm} {h}_{\rm LoS}(f,d) 		+ {h}_{\rm a}(f,d) \\
  \nonumber  \hspace{-0.2cm}&=&\hspace{-0.2cm} \left( \frac{c}{4 \pi f d}\right)  e^{-k(f) \times \frac{d}{2}} \times e^ {j2\pi\frac{d}{\lambda}} \\
    \hspace{-0.2cm}& &\hspace{-0.2cm} + (1-e^{-k(f) d})^{\frac{1}{2}} \Big(\frac{c}{{4\pi f d}}\Big) \times e^ {j2\pi\beta}.
    \label{eq:Hfunc}
\end{eqnarray}

\subsection{MIMO Channel Model and Capacity}

\begin{figure}[h]
\centering
\includegraphics[width=0.45\textwidth]{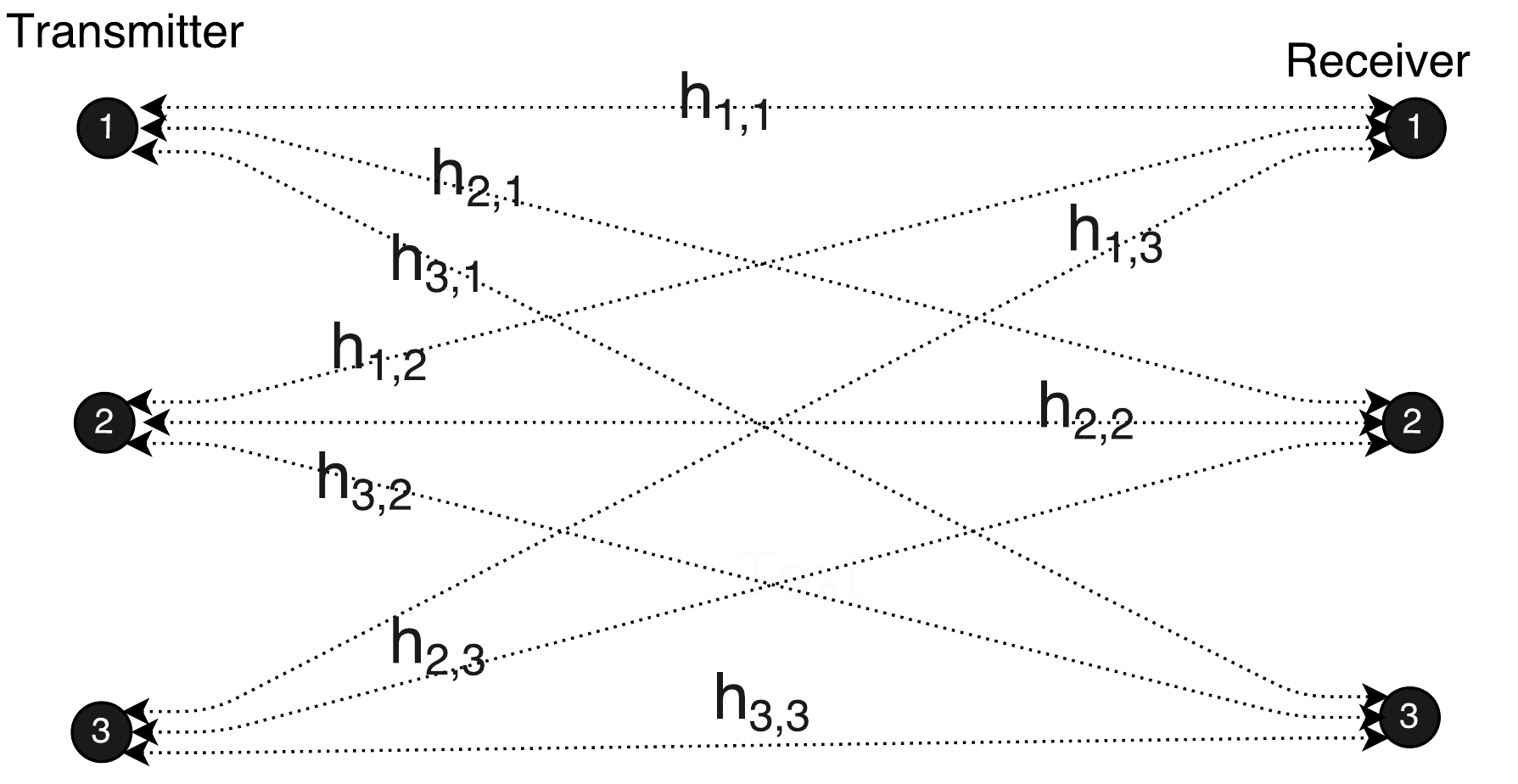}
\caption{A 3x3 MIMO system, the channel transfer coefficient ${{h}_{ij}}$ of each antenna pair between transmitters and receivers is shown. }
\label{fig:Hmatrix}
\end{figure}

In this work, a point to point wireless communications system is considered with multiple antennas where the number of antennas at the receiver and transmitter equals to $n_r$ and $n_t$, respectively. A generic wireless channel is considered where the up-link and down-link directions are outside the scope of the study. 
The received signal vector $\mathbf{y}$ at $n_r$ receiving antennas can be formulated as~\cite{chinani2003MIMOcap}
\begin{equation}
\mathbf{y}=\mathbf{{H}}\mathbf{x}+\mathbf{n},
\label{eq:MIMObase}
\end{equation}
where $\mathbf{x}$ is the transmitted signal vector form $n_t$ transmitting antennas,
and $\mathbf{n}$ is an $n_r\times1$ vector with zero-mean independent noises with variance $\sigma^2$.
The channel matrix $\mathbf{{H}}$ is defined by
\begin{equation}
{\mathbf{H}} \triangleq
\begin{bmatrix}
 &{h}_{11}  &{h}_{12}  & \ldots &{h}_{1n_t} \\
 &{h}_{21}  &{h}_{22}  & \ldots &{h}_{2n_t} \\
 &\vdots  &\vdots  & \ddots & \vdots  \\
 &{h}_{n_r1}  &{h}_{n_r2}  &\ldots &{h}_{n_rn_t}  \\
\end{bmatrix},
\label{eq:Hmatrix}
\end{equation}
where ${{h}_{ij}}$ is a complex value denoting the transfer coefficient between the $i$th receiving antenna and the $j$th transmitting antenna.
Note that ${{h}_{ij}}$ can be obtained from~\eqref{eq:Hfunc} for frequency $f$ and distance $d_{ij}$.
The channel coefficients of a MIMO system consisting of $3$ antennas at both the transmitter and the receiver is presented in \figurename~\ref{fig:Hmatrix}.

For a MIMO system with equally distributed power among the transmitting antennas, the capacity can be written as
\begin{equation}
C = {\rm log}_{2}{\rm det} ({I}_{n_r}+\frac{P}{n_t\sigma^2}\mathbf{{H}}\mathbf{{H}}^\dagger),
\label{eq:MIMOcapBase}
\end{equation}
where $P$ is the total transmit power, $\mathbf{{H}}^\dagger$ is the Hermitian transpose of $\mathbf{{H}}$
and $I$ is the identity matrix~\cite{chinani2003MIMOcap}.

The \emph{singular value decomposition} (SVD) of the channel transfer matrix $\mathbf{{H}}$ can be given by:
\begin{equation}
\centering
\mathbf{{H}} = \mathbf{U\Sigma V^{\dagger}},
\label{eq:SVD1}
\end{equation}
where $\mathbf{U}$ and $\mathbf{V}, $respectively, are $n_r\times n_r$ and $n_t\times n_t$ unitary matrices, $\mathbf{\Sigma}$ is a rectangular diagonal $n_r\times n_t$ matrix and $\mathbf{V^{\dagger}}$ is the Hermitian transpose of $\mathbf{V}$. The non-negative real diagonal elements of matrix $\mathbf{\Sigma}$, $\lambda_1\geq \lambda_2\geq...\geq\lambda_m$, are the ordered singular values of matrix $\mathbf{{H}}$. Hence, the squared singular values $\lambda_i^2$ denote the eigenvalues of the matrix $\mathbf{{H}}\mathbf{{H}}^\dagger$. The SVD decomposition describes an equivalent MIMO system with parallel independent channels where $i_{\text{th}}$ channel is a virtual single-input-single-output (SISO) with a gain $\lambda_{i}$ and the allocated power $P_i$ .
Therefore, the capacity of the MIMO system is equal to the cumulative capacity of the independent SISO channels~\cite{Tse2005MIMObook}
\begin{equation}
C = \sum_{i=1}^{m}{\rm log}_2(1+\frac{P_i\lambda_i^2}{\sigma^2}),
\label{eq:MIMOcapEigWaterFill}
\end{equation}
where $m$ is the number of non-zero $\lambda_i^2$, $m\leq {\rm min}(n_r,n_t)$, which is also called the rank of $\mathbf{{H}}$. $\frac{P_i\lambda_i^2}{\sigma^2}$ is the associated received signal-to-noise ratio (SNR) to each SISO channel. Furthermore, the SNR of the equivalent channel should meet a minimum threshold to be reliably detectable by the receiver. In this paper, we assume 0\,dB as the SNR threshold.

\subsection{SVD-Based Precoding }
\label{sec:precod}

Precoding technology assigns different precoder weights to independent data streams transmitted by multiple antennas, aiming to enhance the transmission rate or the transmission reliability by exploiting the spatial dimension~\cite{park2009new}. We assume that the transmitter has $z$ independent data streams to transmit simultaneously, denoted by $\mathbf{s}=[s_1,...,s_z]^T$, $z\leq{\rm min}(n_r,n_t)$. With the linear precoding scheme, a precoding matrix $\mathbf{W}$ is used to generate the precoded signal vector $\mathbf{x}$,

\begin{equation}
\centering
\mathbf{x}=\mathbf{W}\mathbf{s}.
\label{eq:precod1}
\end{equation}

We assume the precoder matrix is orthonormal i.e., $\mathbf{W^\dagger W}=\mathbf{I}_z$ \cite{park2009new}.

The optimal linear precoder in term of capacity can be obtained by choosing $\mathbf{W}=\mathbf{V}_z\mathbf{Q}$, where the $\mathbf{V}_z$ is an $n_t\times z$ matrix constructed by the first $z$ columns of $\mathbf{V}$ in~\eqref{eq:SVD1} and $\mathbf{Q}$ is a diagonal matrix for power allocation~\cite{goldsmith2005wireless,park2009new}. If $P$ is the total power constraint, 

\begin{equation}
\centering
\mathbf{Q}=diag\{\sqrt{P_i},..., \sqrt{P_z}\},\quad \text{where} \quad \left(\sum_{i=1}^{z}P_i\right)\leq P.
\label{eq:Qmatrix}
\end{equation} 

\subsubsection{Optimal Beamforming}
\label{sec:OBF-precod}

In the absence of multipath channel, a single stream beamforming (BF) is used to focus energy in one direction and improve the SNR. In this approach, the same copy of data is sent to all transmitter antennas to form a directional beam and exploit the high single power gain~\cite{Sibille2011LTEbook}. Therefore, $z=1$ and the optimal precoder vector is

\begin{equation}
\centering
\mathbf{W}=\sqrt{P}\mathbf{V}_1,
\end{equation}
where $\mathbf{V}_1$ is the first column of $\mathbf{V}$~\cite{bouhlel2015transmit}. This technique also is known as \emph{Closed-loop rank-1 precoding} and is optimal for the LoS channel where the rank of the channel transfer matrix is one~\cite{Sibille2011LTEbook}.

\subsubsection{Closed-Loop Spatial Multiplexing}
\label{sec:MPcSI-precod}

As mentioned above, multiplexing aims to exploit spatial multiplexing to increase channel throughput. Hence, it is optimal when exploiting all the parallel sub-channels to maximize the channel capacity. But the number of parallel channels depends on the channel rank, the so-called degree-of-freedom, i.e., the number of non-zero singular values in $\mathbf{\Sigma}$. The channel is full rank if the channel matrix $\mathbf{{H}}$ is sufficiently random. In presence of Channel State Information (CSI) at the transmitter, the optimal multiplexing is to send an independent data stream on each sub-channel which is known as \emph{Closed-Loop Spatial Multiplexing}. In other words, if we consider a full rank channel, $z={\rm min}(n_r,n_t)$ and

\begin{equation}
\centering
\mathbf{W}=\mathbf{VQ}.
\end{equation}

The optimal power allocation $\mathbf{Q}$ can be obtained by the water-filling scheme. In this way, the equivalent virtual SISO channel with a larger channel gain, $\lambda_i$, will be allocated with a larger transmit power, $P_i$. That is, 

\begin{equation}
P_i = \left(\mu - \frac{\sigma^2}{\lambda_i^2}\right),
\label{eq:waterFill}
\end{equation}
where $\mu$ is chosen to satisfy the overall power constraint.

Hence, the transmitter needs to know all the elements of the matrices $\mathbf{V}$ and $\mathbf{\Sigma}$ for precoding and adopting the water-filling. 

Note that for a pure LoS channel with rank 1, there is only one independent sub-channel. Therefore, both multiplexing and beamforming use the precoding matrix, $\mathbf{W}=\sqrt{P}\mathbf{V}_1$, which results in the same capacity. 

\subsubsection{Open-Loop Spatial Multiplexing}
\label{sec:MPcSI-precod}

When the channel information is unknown, an Open-loop spatial multiplexing scheme can be used to maximize the spatial multiplexing capacity~\cite{goldsmith2005wireless,Sibille2011LTEbook}, where the equal power allocation and the identity matrix, $\mathbf{I_{n_t}}$, are used to adapt the precoding matrix. That is,
\begin{equation}
\mathbf{W}=\sqrt{P}\mathbf{I}_{n_t}.
\label{eq:blind}
\end{equation}

Also, due to the unknown channel information, data can be decoded using maximum-ratio combining (MRC) method~\cite{Tse2005MIMObook}.


\section{MIMO Capacity with Molecular Re-radiation Modeled as Scattering}
\label{sec:analysis}

In this section, we analyze MIMO capacity when molecular re-radiation is assumed to be scattering. To quantitatively characterize the scattering richness of the channel, we firstly decompose and normalize the channel transfer function, $\mathbf{H}$, as

\begin{equation}
\begin{aligned}
\hat{\mathbf{H}}(f,d) = \sqrt{\frac{K}{K+1}}\hat{\mathbf{H}}_{\rm LoS}(f,d) + \sqrt{\frac{1}{K+1}}\hat{\mathbf{H}}_{\rm a}(f,d),
\label{eq:HfuncM2}
\end{aligned}
\end{equation}
where $\hat{\mathbf{H}}$, $\hat{\mathbf{H}}_{\rm LoS}$ and $\hat{\mathbf{H}}_{\rm a}$ are normalized matrices of ${\mathbf{H}}$, ${\mathbf{H}}_{\rm LoS}$ and ${\mathbf{H}}_{\rm a}$, respectively. Because of the uniformly distributed random phase of the large number of re-radiated signals, the elements of $\hat{\mathbf{H}}_{\rm a}$ are approximately independent and identically
distributed (i.i.d) complex Gaussian random variables~\cite{asztely1998effects} with zero
mean and unit magnitude variance. Denoted by $K$, the ratio of powers of the LoS signal and the re-radiated components under the assumption that the channel distance is much larger than the antenna array size, can be obtained by

\begin{equation}
\begin{aligned}
K = \frac{P_{\rm r,LoS}(f,d)}{P_{\rm r,a}(f,d)} = \frac{e^{-k(f) d}}{1-e^{-k(f) d}}.
\label{eq:k-factor}
\end{aligned}
\end{equation}

\begin{figure}[]
	\begin{center}
		\includegraphics[width=0.5\textwidth ,clip=true, trim=0 0 0 0]{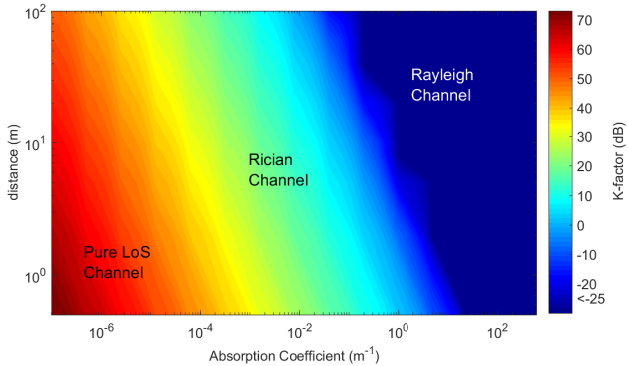}
		\caption{Rician K-factor varies with the absorption coefficient and distance.}	
		\label{fig:K-factor_analys}
	\end{center}
	\vspace{-0.5cm}
\end{figure}

This is also known as the Rician K-factor in Rician fading model. Equivalently, K-factor represents the channel richness in terms of scattering and multipath rays. Equation~\eqref{eq:k-factor} shows that $K$ is a decreasing function of both the absorption coefficient $k(f)$ and the distance between the transmitter and receiver $d$. \figurename~\ref{fig:K-factor_analys} illustrates how K-factor changes with the absorption coefficient in the atmosphere for a distance of 1-100m. The MIMO capacity in relation to the Rician K-factor has been well studied in the literature~\cite{Jayaweera2002Rician},~\cite{farrokhi2001link} and~\cite{lebrun2006mimo}.

Obtaining a closed-form expression of the MIMO capacity as a function of $k(f)$ is very complicated. Hence, we state the lower and upper bounds of the expected capacity of the MIMO channel as a function of $k(f)$ in two lemmas. Lower bound expression is stated as Lemma 1 based on~\cite{lebrun2006mimo} where authors showed the lower bound of the expected capacity of a Rician channel.

\begin{widetext}
\newtheorem{lem}{Lemma}

\begin{lem}
 When the transmitter does not have CSI and the power is allocated equally to all transmitter antennas, the capacity lower bound of a Rician channel is the capacity contributed by the NLoS component:
\begin{equation}
\begin{aligned}
E(C(\hat{\mathbf{H}}),\rho)\geq E(C({\hat{\mathbf{H}}}_a),\sqrt{\frac{1}{K+1}}\rho),
\label{eq:lower-band1}
\end{aligned}
\end{equation}
\begin{equation}
\begin{aligned}
 \implies E(C(\hat{\mathbf{H}}),\rho)\geq E(C({\hat{\mathbf{H}}}_a),\sqrt{1-e^{-k(f) d}}\rho),
\label{eq:lower-band2}
\end{aligned}
\end{equation}
where $E(.)$ denotes the expectation, $\rho$ is the received SNR for equivalent single channel and $E(C(\hat{\mathbf{H}},\rho)$ is the average capacity of a channel with normalized channel transfer matrix $\hat{\mathbf{H}}$ and a reception SNR $\rho$. It is clear that the lower bound is an increasing function of the absorption coefficient.

\begin{proof}
	See \cite{lebrun2006mimo}.
\end{proof}
\end{lem}

 On the other hand, the capacity upper bound of a Rician MIMO channel with perfect CSI at receiver and transmitter where the power is optimally distributed among antennas with a \emph{water-filling} algorithm is also presented in~\cite{Jayaweera2002Rician}:
\begin{equation}
\begin{aligned}
E(C(\hat{\mathbf{H}}), \rho)\leq &{\rm log}_2 \left(1+\frac{n_r(1+n_tK)}{K+1}\Big({\rm min}\left\{\frac{\rho}{n_t},\frac{K(1+K)}{n_r(1+n_tK)}\right\}n_t+\left[\frac{\rho}{n_t}-\frac{K(1+K)}{n_r(1+n_tK)}\right]^{+}\Big)\right)+ \\
&(n_t-1){\rm log}_2 \left(1+\frac{n_r}{1+K}\left[\frac{\rho}{n_t}-\frac{K(1+K)}{n_r(1+n_tK)}\right]^{+} \right),
\label{eq:upper-band1}
\end{aligned}
\end{equation}
\normalsize
where $[x]^{+} = {\rm max}(x,0)$. 

To obtain a simpler expression, let us consider a specific scenario where $n_r=n_t=n$:

 \begin{lem} 
 When there is perfect CSI at receiver and transmitter and the power is allocated optimally to transmitter antennas, the capacity upper bound of a Rician channel is
 \begin{equation}
\begin{aligned}
E(C(\hat{\mathbf{H}}),\rho)\leq &{\rm log}_2 \left(1+(1+(n-1)e^{-k(f) d})\Big({\rm min}\left\{\rho ,\frac{\frac{e^{-k(f) d}}{1-e^{-k(f) d}}}{(1+(n-1)e^{-k(f) d})}\right\}n+\left[\rho -\frac{\frac{e^{-k(f) d}}{1-e^{-k(f) d}}}{(1+(n-1)e^{-k(f) d})}\right]^{+}\Big)\right) \\
&+(n-1){\rm log}_2 \left(1+(1-e^{-k(f) d})\left[\rho,\frac{\frac{e^{-k(f) d}}{1-e^{-k(f) d}}}{(1+(n-1)e^{-k(f) d})}\right]^{+} \right).
\label{eq:upper-band2}
\end{aligned}
\end{equation}

\normalsize
\begin{proof}
	See \cite{Jayaweera2002Rician}.
\end{proof}

 \end{lem}

\end{widetext}

We can see two extreme cases: the low absorption $k(f)=0$ and the high absorption $k(f)=\infty$. Hence, the capacity of a Rician channel for MIMO will limit to
\begin{equation}
\begin{aligned}
\lim_{k(f) \to \infty }C = n\rm{log}(1+\rho),
\label{eq:limit-uband}
\end{aligned}
\end{equation}
\begin{equation}
\begin{aligned}
\lim_{k(f) \to 0 }C = \rm{log}(1+n^{2}\rho).
\label{eq:limit-lband}
\end{aligned}
\end{equation}

Note that from~\eqref{eq:k-factor}, $k(f)=\infty$ means an extremely high re-radiation channel, which provides a pure Rayleigh channel. In contrast, $k(f)=0$ implies no re-radiation, i.e., a pure LoS channel.

\section{Simulation}
\label{sec:simulSetup}
In this section, simulation results of MIMO in the presence of molecular absorption and re-radiation are presented.

\subsection{Simulation Set-up}
\label{sec:geometry}

\begin{figure}[h]
\centering
\includegraphics[width=0.4\textwidth ,clip=true, trim=200 0 100 0]{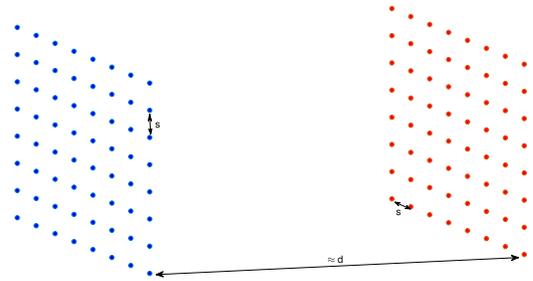}
\caption{Illustration of a MIMO system with uniform square arrays.}
\label{fig:geometry}
\end{figure}

\begin{table*}[h]
\centering
\caption{Simulation parameters}
\label{tab:simParam}
\begin{tabular}{|l|l|}
\hline
Transmitter and receiver distance ($d$) & $1\sim50$\,m                       \\
Inter-element spacing ($s$)             & $0.5 \lambda$ (wave length)        \\
Transmitter array angle           & random \\
Receiver array angle                    & random \\
Number of antennas on each side ($n$)     & $64$ in mmWave, $225$ in terahertz \\
SNR (in fixed SNR configurations) & $5,15$\,dB \\ 
Non-molecular noise & $-80$ dBm\\ 
Transmit power & $150$\,mW in mmWave and $1$\,mW, $10$\,mW in terahertz \\ \hline
\end{tabular}
\end{table*}

To evaluate the performance impact of the molecular absorption on the MIMO capacity in the mmWave/terahertz band,
we consider an $n\times n$ MIMO system with two square uniform arrays as illustrated in \figurename~\ref{fig:geometry}. The inter-element spacing, $s$, is equal to half of the wavelength while the channel distance, $d$, represents the distance between the transmitter and the receiver. All simulations parameters are defined in Table~\ref{tab:simParam}. Some of these parameter values are varied to achieve a more comprehensive evaluation of the MIMO system under different conditions. Given that molecular re-radiation produces random phases, a large number (5000) of test cases are evaluated and averaged to obtain the result for each parameter value. 

The absorption coefficients are obtained from HITRAN~\cite{Rothman2012Database} for some predefined standard gas mixtures of the atmosphere at sea level, which is accessible online
\footnote{http://hitran.iao.ru/gasmixture/simlaunch}. The related information of the predefined gas mixtures is shown in Table~\ref{tab:gasMix}.
Recall that the oxygen and water molecules are the main absorption players in the normal atmosphere at mmWave/terahertz bands~\cite{hoseini2016diurnal}. While the oxygen ratio is invariant, the amount of water molecules in the air is variable. 
We use the highest and lowest water ratio in Table~\ref{tab:gasMix},
i.e., the \emph{"USA model, high latitude, winter"} and \emph{"USA model, tropics"}.
The corresponding absorption coefficients are shown in \figurename~\ref{fig:kflogmmWave} and \figurename~\ref{fig:kflogTHz} for the mmWave and terahertz band, respectively, which are calculated using the ambient temperature of $273$\,K and the sea level pressure of $1$\,atm.

\begin{table*}[]
	\centering
	\caption{Atmosphere standard gas mixture ratio in percentage for different climates at sea level \cite{Rothman2012Database}}

	\label{tab:gasMix}
	\begin{tabular}{|l|l|l|l|l|l|l|l|l|}
		\hline
		USA model: & \multicolumn{1}{c|} {H2O} & \multicolumn{1}{c|} {CO2} &  \multicolumn{1}{c|} {O3} &  \multicolumn{1}{c|} {N2O} &  \multicolumn{1}{c|} {CO} & \multicolumn{1}{c|} {CH4} & \multicolumn{1}{c|} {O2} &  \multicolumn{1}{c|} {N2} \\ \hline
		mean latitude, summer & 1.860000 &  0.033000 & 0.000003 & 0.000032 & 0.000015 &  0.000170 & 20.900001 &  77.206000 \\ \hline
		mean latitude, winter & 0.432000 &  0.033000 & 0.000003 & 0.000032 & 0.000015 &  0.000170 & 20.900001 &  78.634779 \\ \hline
		high latitude, summer & 1.190000 &  0.033000 & 0.000002 & 0.000031 & 0.000015 &  0.000170 & 20.900001 &  77.876781 \\ \hline
		high latitude, winter & 0.141000 &  0.033000 & 0.000002 & 0.000032 & 0.000015  & 0.000170 & 20.900001 &  78.925780 \\ \hline
		tropics & 2.590000 &  0.033000 & 0.000003 & 0.000032 & 0.000015  & 0.000170 & 20.900001 &  76.476779 \\ \hline
	\end{tabular}
\end{table*}

\subsection{Impact of Absorption Coefficient on MIMO Capacity}\label{sec:ResaultsKF}

\begin{figure}[]
\centering
\includegraphics[width=0.5\textwidth]{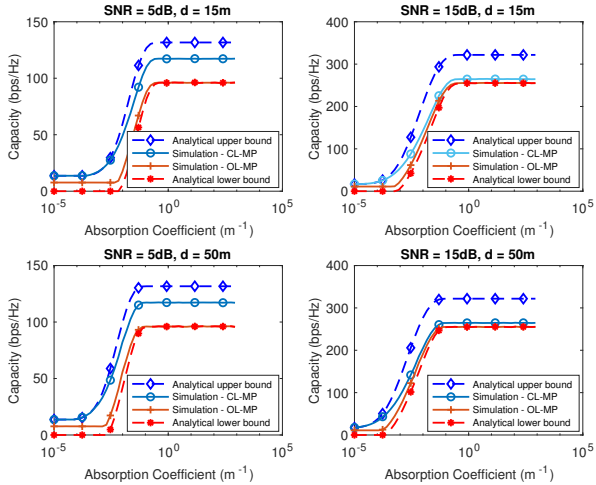}
\caption{Analytical bounds and simulation results of 64x64 MIMO capacity versus absorption coefficient for different SNR and distance. The re-radiation is assumed as scattering.}
\label{fig:Anlytical-simulation}
\end{figure}

\begin{figure}[]
\centering
\includegraphics[width=0.5\textwidth ,clip=true, trim=0 0 0 5]{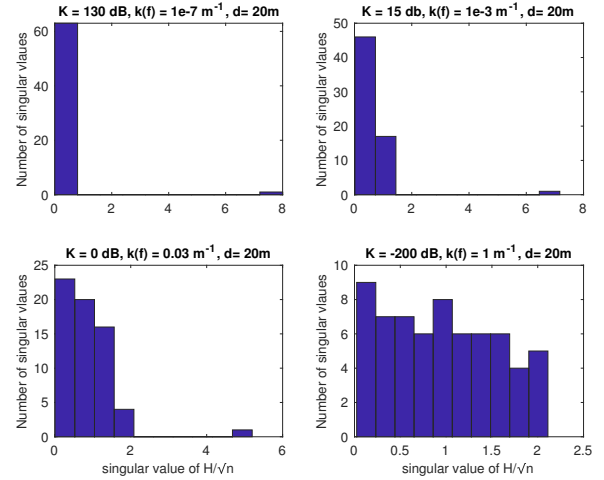}
\caption{Empirical distribution of singular values of matrix $\frac{H}{\sqrt{n}}$ for different K-factor. For $K \to -\infty $ dB, it converges to the quarter circle law.}
\label{fig:singularValueHist}
\end{figure}

In this subsection, MIMO performance is evaluated with the re-radiation modeled as scattering. A 64x64 MIMO system at 60\,GHz is investigated with a realistic absorption coefficient range between 10\textsuperscript{-5}$\sim$10\textsuperscript{+3}. 
The uniform and the water-filling power allocations are plotted and labeled in \figurename~\ref{fig:Anlytical-simulation} as \emph{OL-MP} and \emph{CL-MP}, respectively. 

From \figurename~\ref{fig:Anlytical-simulation}, it can be observed that the capacity of MIMO lies between the theoretical bounds, which verifies the lemmas proposed in Section~\ref{sec:analysis}. An interesting observation is that the MIMO capacity increases with higher absorption coefficients. For example, under 5\,dB SNR, we obtain a capacity of only 9.5\,bps/Hz when the absorption is very low, i.e., the MIMO experiences a pure LoS channel. In contrast, we obtain 100\,bps/Hz with uniform power allocation and 130\,bps/Hz with water-filling when the absorption coefficient is high. This result indicates that improved MIMO capacity can be expected for frequency windows where the absorption coefficient of the channel peaks. Finally, we notice that the capacity curve against the absorption coefficient starts to increase sooner for larger distances, which can be explained by the fact that a physically longer channel would contain more molecules to create more intense scattering. This phenomenon can also be seen in \figurename~\ref{fig:K-factor_analys}, where a longer distance results in a smaller Rician K-factor. However, a longer transmission distance would also lead to a larger path loss, which would eventually have a negative effect on MIMO capacity for the long-distance communications. The impact of distance on the MIMO capacity will be evaluated in more detail in Section~\ref{sec:MIMOmmWave}.

From the other perspective, the re-radiation changes the channel transfer matrix from a deterministic one to a random matrix. When the channel matrix is deterministic, it has a \emph{degree-of-freedom} equal to one where the first singular value is significant and the others are almost zero. The molecular re-radiation changes the singular value distribution of the channel matrix. We show the empirical distribution of singular value with various absorption coefficient in \figurename~\ref{fig:singularValueHist}. It can be seen that the distribution converges to the well-known quarter-circle law,~\cite{tse2005fundamentals}, with a large absorption coefficient.

\subsection{Impact of Absorption Coefficient on MIMO Channel Characteristics}\label{sec:MIMOmmWave}

\begin{figure}
    \centering
    \includegraphics[width=0.5\textwidth ,clip=true, trim=0 0 0 10]{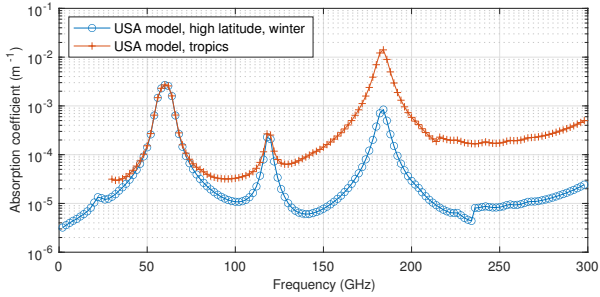}
    \caption{Absorption curves in different climates (temperature=$273~K$, pressure=$1~atm$). Absorption peaks are exhibited at resonance frequencies of oxygen and water molecules.}
    \label{fig:kflogmmWave}
\end{figure}

\begin{figure}[]
	\center
	\subfloat[Channel rank]{\label{fig:rankmmWave}
		\includegraphics[width=0.50\textwidth ,clip=true, trim=0 0 0 10]{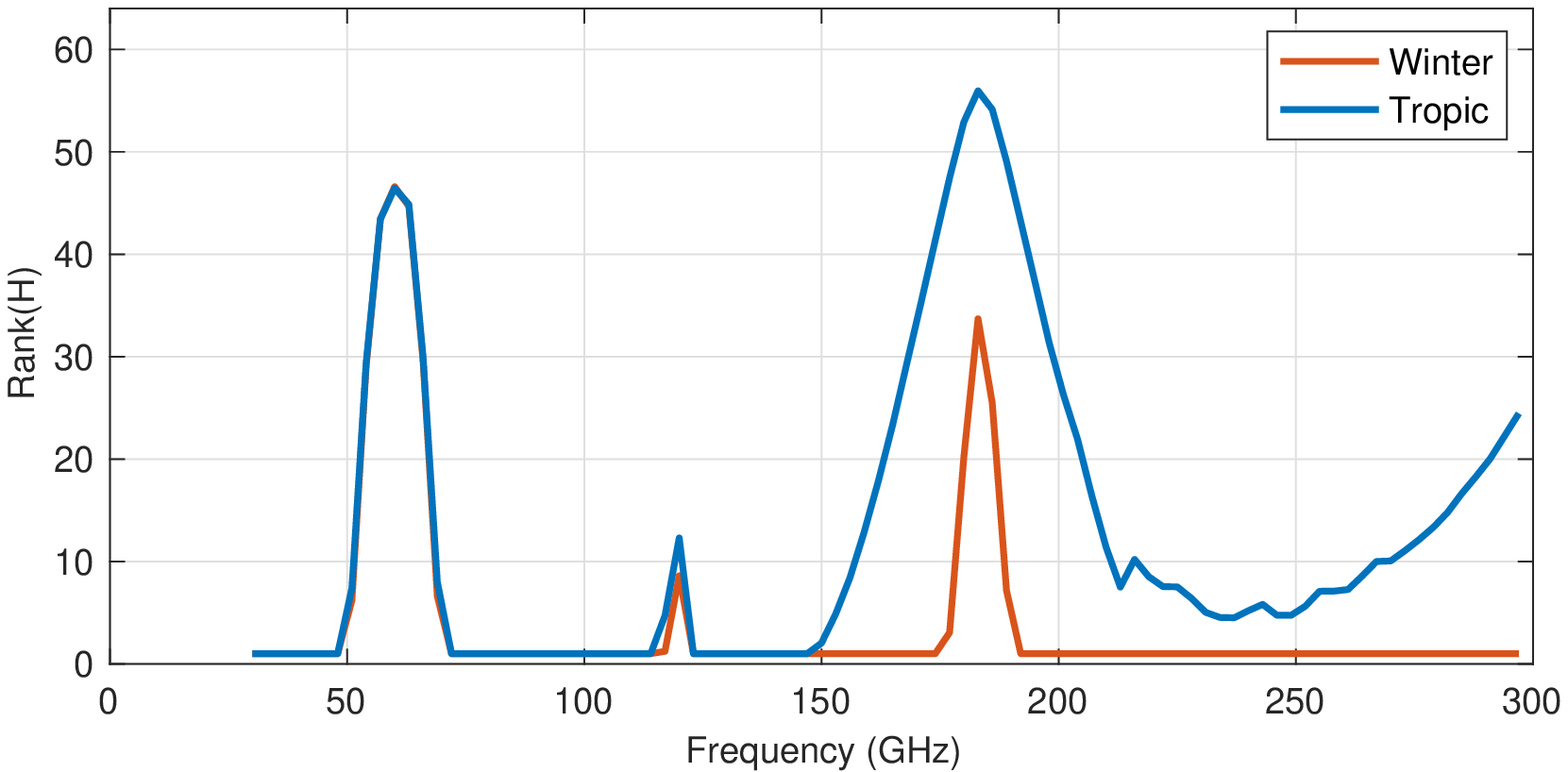}}\vspace{-0.1em}
	\subfloat[K-factor]{\label{fig:KmmWave}
		\includegraphics[width=0.50\textwidth ,clip=true, trim=0 0 0 10]{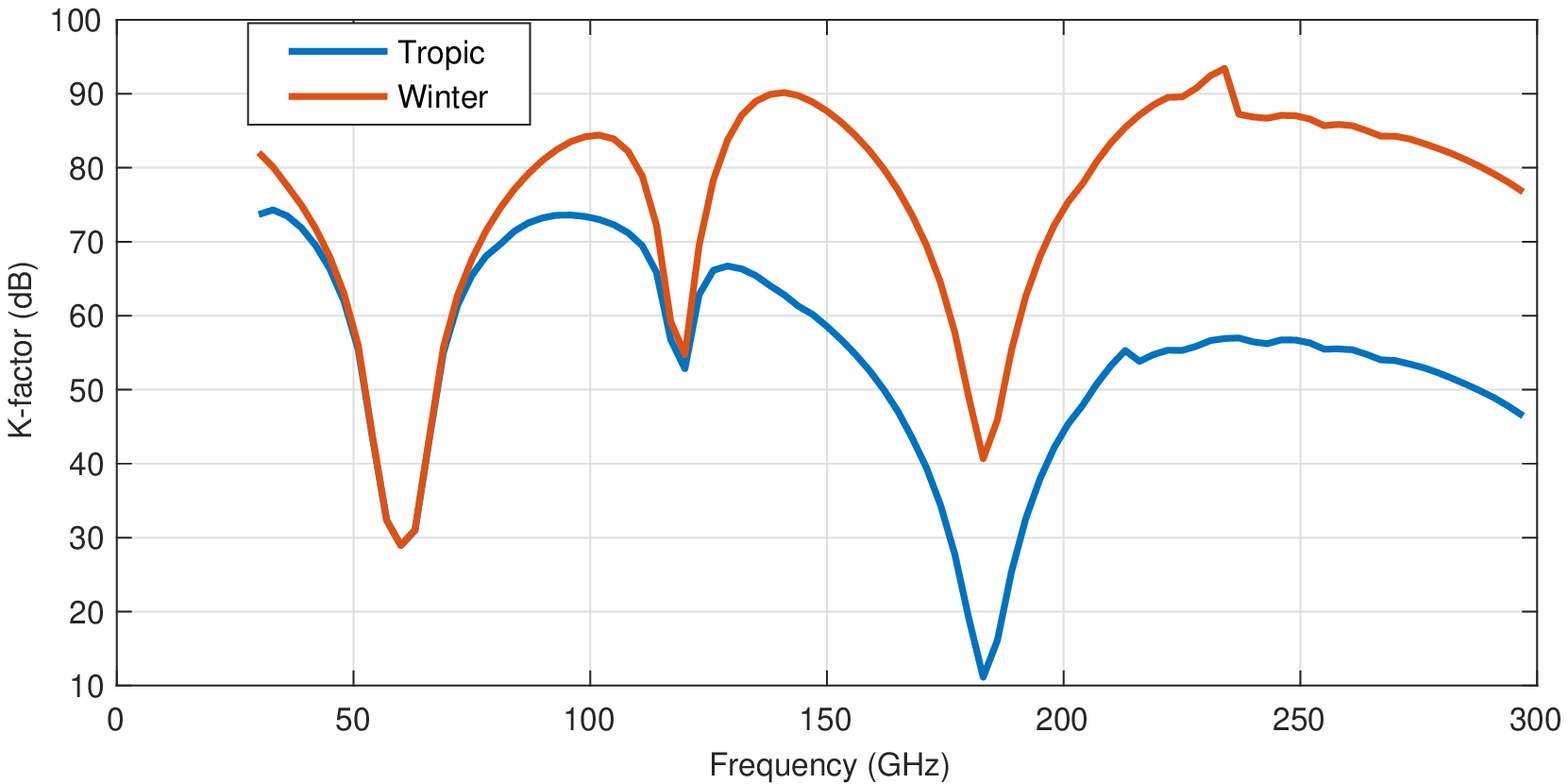}}\vspace{-0.1em}
	\subfloat[Condition number]{\label{fig:CondNmmWave}
		\includegraphics[width=0.50\textwidth ,clip=true, trim=0 0 0 5]{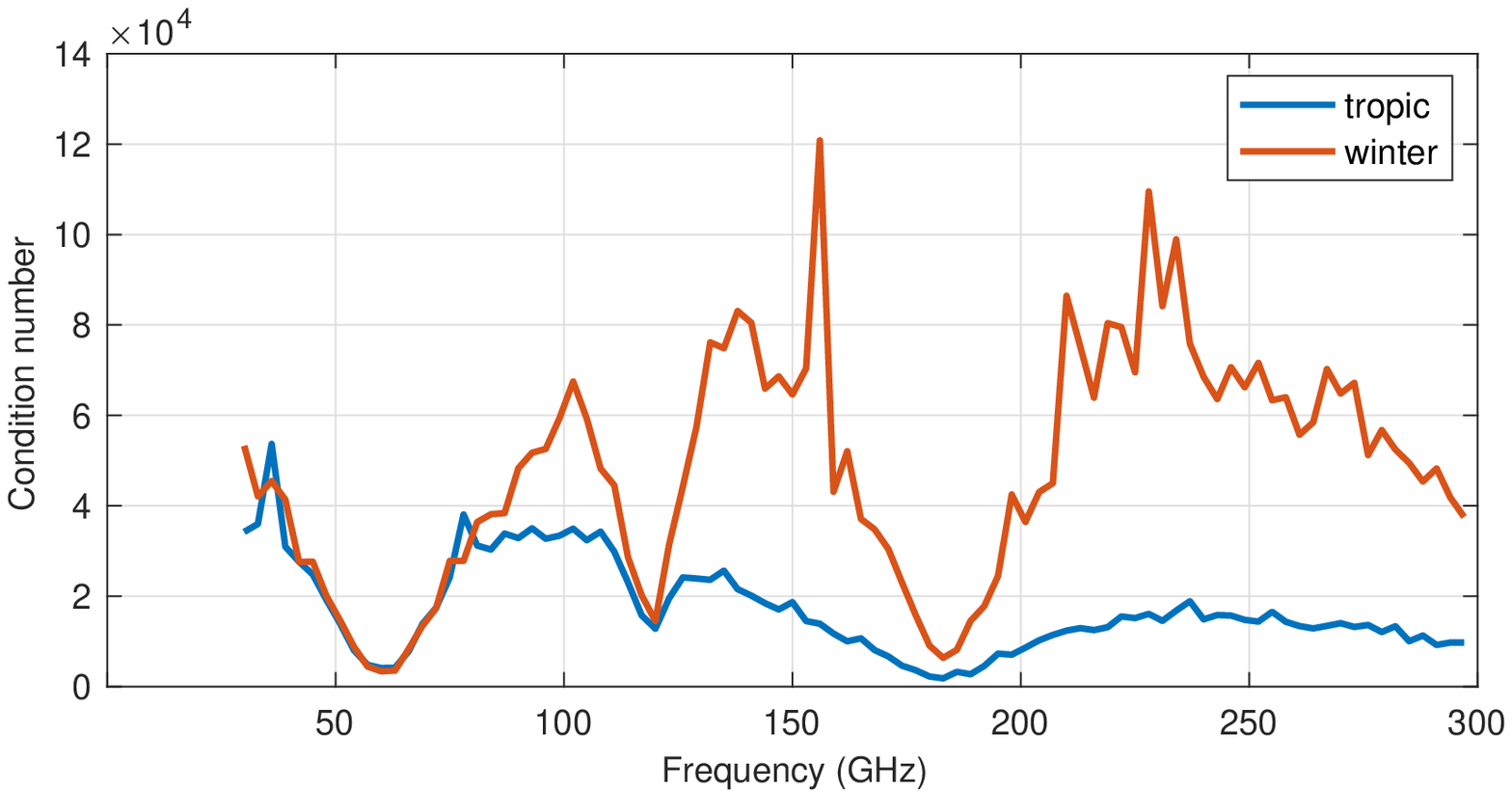}} \\
	\caption{Channel characteristics of a 64x64 MIMO channel over mmWave band. Channel rank, K-factor and condition number are affected by absorption peaks at 60, 120 and 180\,GHz. }
	\label{fig:mmWaveResult0}
\end{figure}

In \figurename~\ref{fig:kflogmmWave},
the channel absorption coefficient against frequency is plotted for the mmWave band for high latitude winter and tropic atmosphere. It is observed that the absorption peaks at 60\,GHz and 120\,GHz, which are the resonance frequencies of oxygen molecules. Absorption also peaks at 180\,GHz, which is the resonance frequency of water molecules. For the two climates, the significant difference in the absorption coefficient at 180\,GHz is due to the significant difference in their water contents in the atmosphere. This result clearly shows that the oxygen and water resonance frequencies are the main frequencies in the high-frequency bands where we can expect high absorption and re-radiation.  

Next, we investigate the impact of the absorption peaks on MIMO channel characteristics. In particular, we are interested to observe how the channel rank, K-factor and condition number are affected by absorption peaks. 

For a 64x64 MIMO, \figurename~\ref{fig:mmWaveResult0} plots the channel rank, K-factor and condition number over the mmWave band for high latitude winter (low water content) and tropic atmosphere (high water content). We know that the physical MIMO channel can be modeled with $m$ equivalent parallel channels with gains given by the singular values $\lambda_i$ where $m={\rm min}(n_r,n_t)$ and $i$= $1$, $2$, ...,$m$. The channel rank is defined by the number of equivalent channels that provide decent SNRs greater than a threshold, which is assumed here as the receiver sensitivity of $0$ dB. \figurename~\ref{fig:rankmmWave} shows that the channel rank peaks at absorption peaks. However, the absorption is not high enough in the mmWave band to yield a full rank channel (rank is still below 64).  \figurename~\ref{fig:KmmWave} shows that the K-factor of the mmWave band drops at absorption peaks, which indicates that there are stronger multi-path signal components resulting from stronger molecular re-radiation at those peaks. For example, we see that the K-factor is about 10\,dB at 180 GHz for the tropic atmosphere, which indicates that the received signal power from the NLoS is as strong as that received from the LoS. Finally, we find (see \figurename~\ref{fig:CondNmmWave}) that the condition numbers at 60\,GHz and 180\,GHz decrease to almost 1, which indicates that the distribution of the eigenvalues is significantly improved at those peaks.

\subsection{MIMO Capacity vs. Re-radiation as Noise/Scattering}

\begin{figure}[t]
\centering

 \subfloat[Constant transmit power = 500\,mW]{\label{fig:cap2constantPow}
	\includegraphics[width=0.5\textwidth ,clip=true, trim=18 0 -10 15]{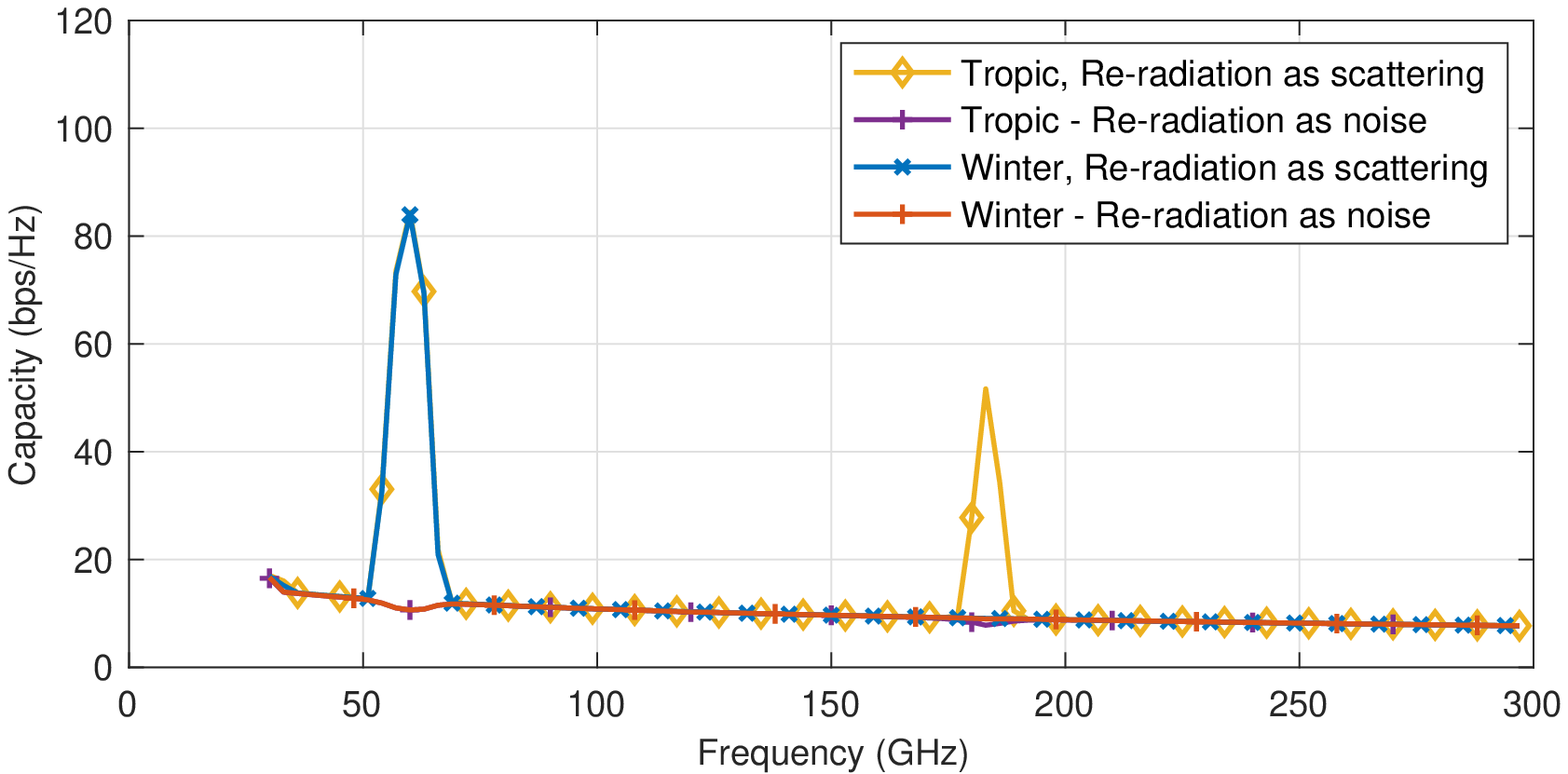}}\vspace{-0.2em}
 \subfloat[Constant SNR = 15\,dB]{\label{fig:capmmWave}
	\includegraphics[width=0.48\textwidth ,clip=true, trim=0 0 0 0]{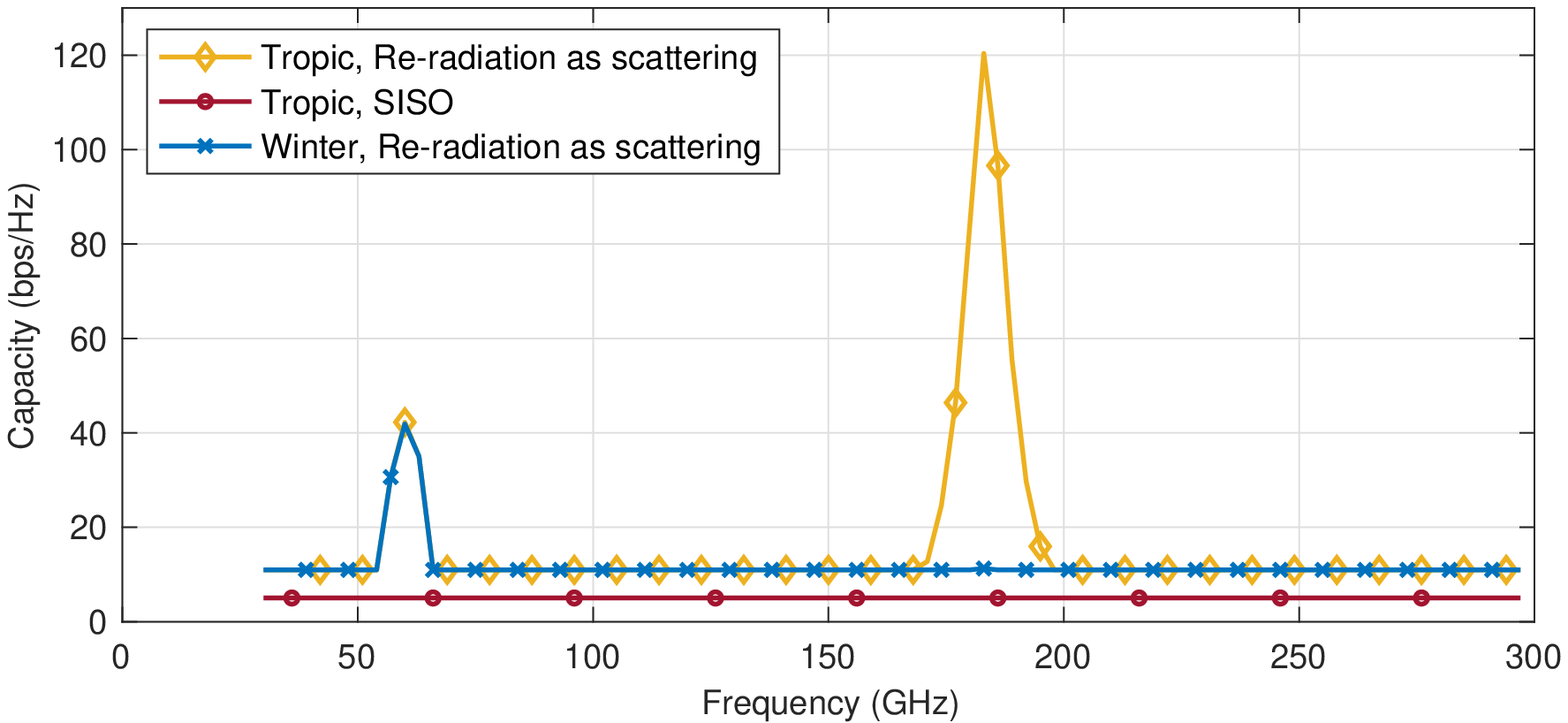}}
\caption{64x64 MIMO capacity for different atmospheres over mmWave band using CL-MP and SISO. Communication distance is 10\,m.}
\label{fig:mmWaveResaults1}
\end{figure}

The purpose of the evaluations in this subsection is to compare MIMO capacity for two different situations: (1) when re-radiation is considered as noise versus (2) when it is considered as a scattered signal. \figurename~\ref{fig:cap2constantPow} compares the capacity of 64x64 MIMO channel in the mmWave band when the re-radiation is assumed as noise versus scattering. The transmit power is constant over the entire band. As one can observe, the capacity, when the re-radiation is considered as noise, is almost flat over the entire mmWave band for both tropic and winter atmospheres and just a very small decrease can be seen in high absorption frequencies which is due to both molecular attenuation and noise. In more details, the molecular noise is dominated by thermal noise and it can be seen the capacity is not frequency selective, which corroborates the common understanding up to now.

In \figurename~\ref{fig:capmmWave}, the received SNR is assumed to be a constant value of 15\,dB for the entire mmWave band to investigate the effect of absorption intensity in an equal SNR condition. It can be seen that,
when the re-radiation is considered as a scattered copy of the signal,
results show a significant capacity increase in particular frequencies.
For example,
the capacity is improved by around 150\,\% at 60\,GHz for both tropic and winter atmospheres.
Moreover,
more water molecules exist in the tropic atmosphere which leads to a considerable capacity boost at 180\,GHz in the tropic atmosphere in comparison with those of the winter atmosphere. Note that the oxygen abortion at 120\,GHz is not strong enough to affect the channel.

\begin{figure}[]
    \begin{center}
            \subfloat[distance = 5 m]{\label{fig:cap5m150mW}
\includegraphics[width=0.5\textwidth, clip=true, , trim= 0 0 0 15]{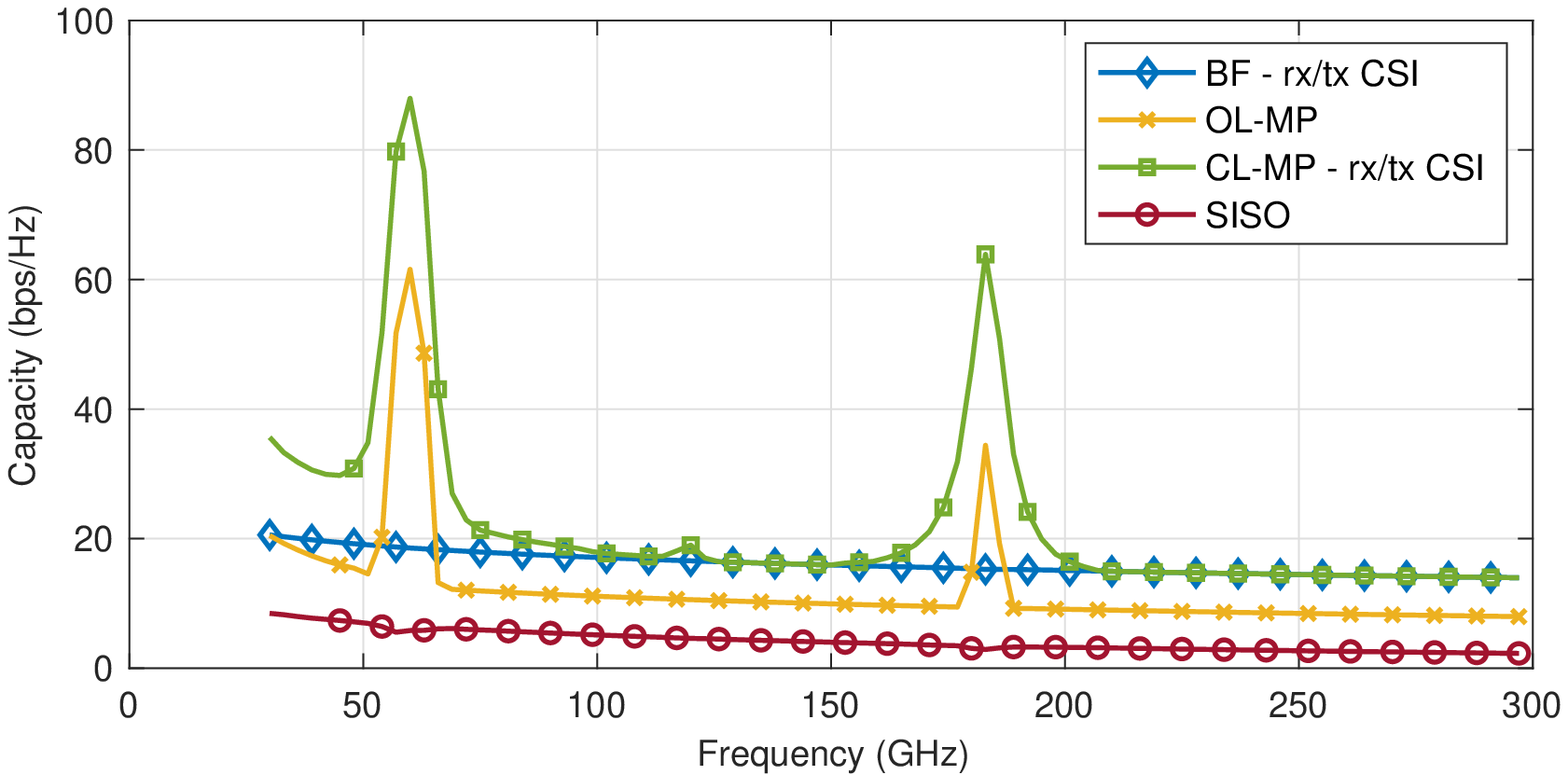}} \vspace{-0.2em}
        \subfloat[distance = 20 m]{\label{fig:cap15m150mW}
\includegraphics[width=0.5\textwidth ,clip=true, trim=0 0 0 0]{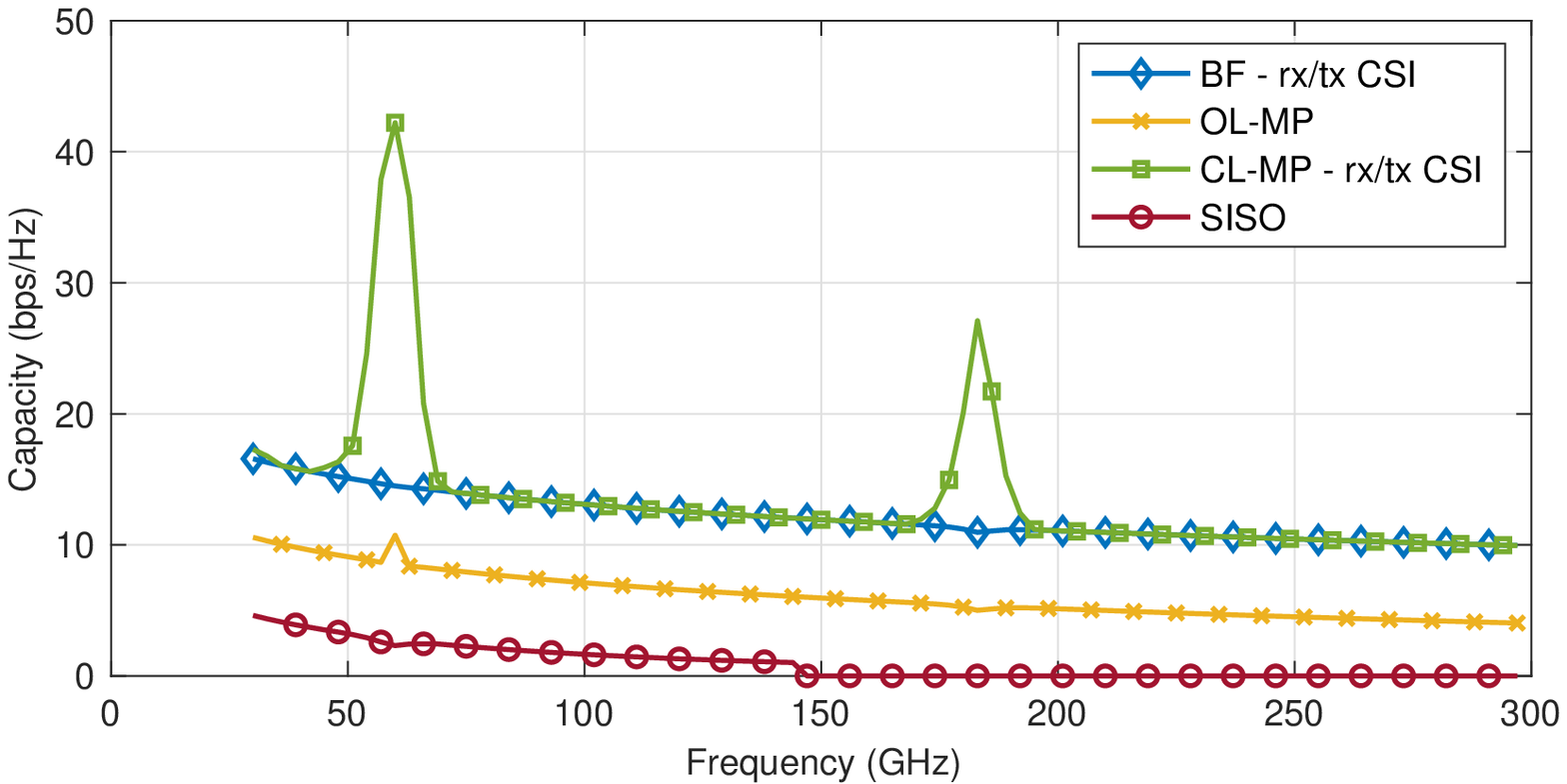}} \vspace{-0.2em}
        \subfloat[distance = 35 m]{\label{fig:cap35m150mW}
\includegraphics[width=0.5\textwidth ,clip=true, trim=0 0 0 0]{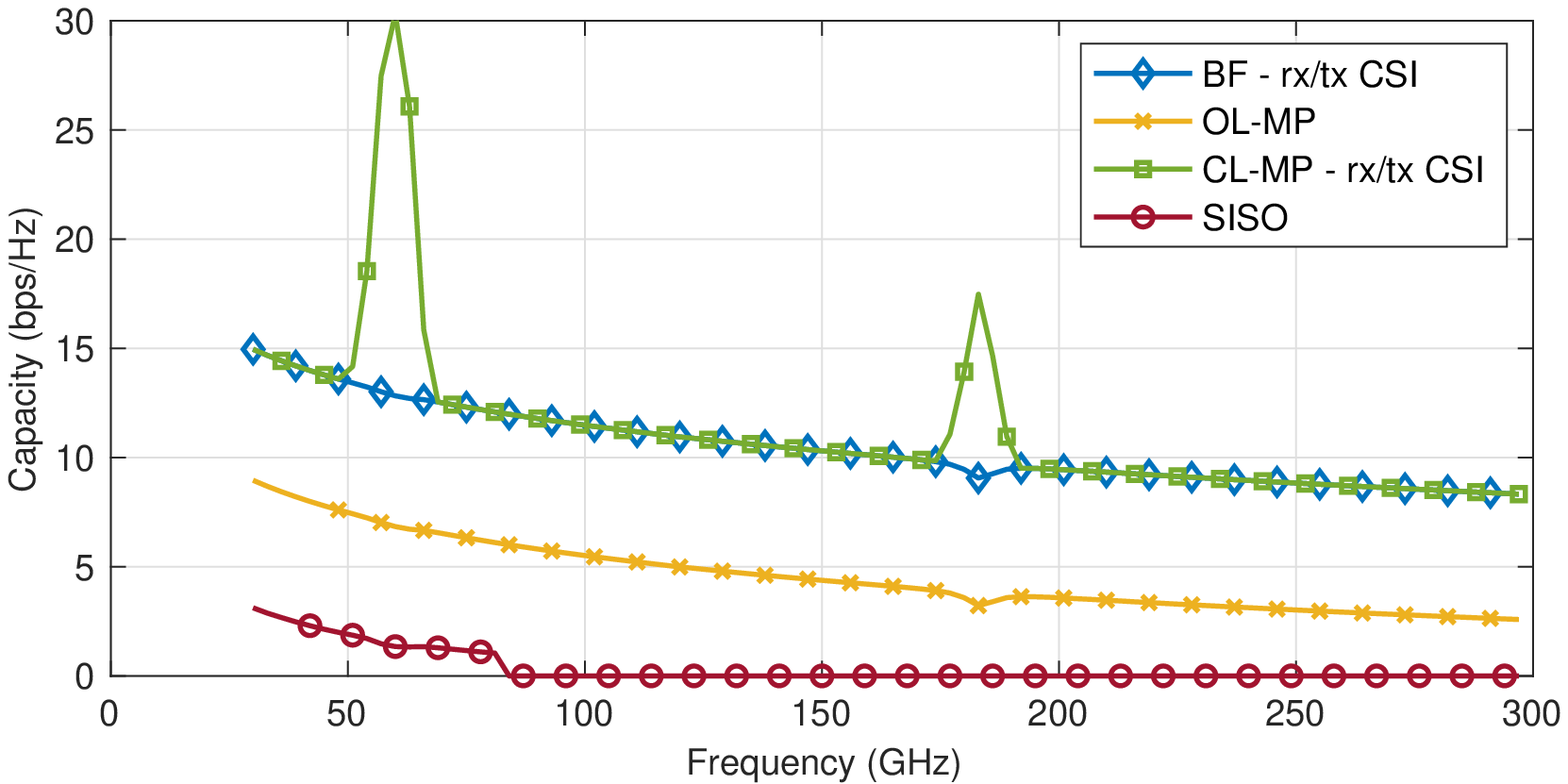}}  \vspace{-0.2em}
        \subfloat[distance = 50 m]{\label{fig:cap50m150mW}
\includegraphics[width=0.5\textwidth ,clip=true, trim=0 0 0 0]{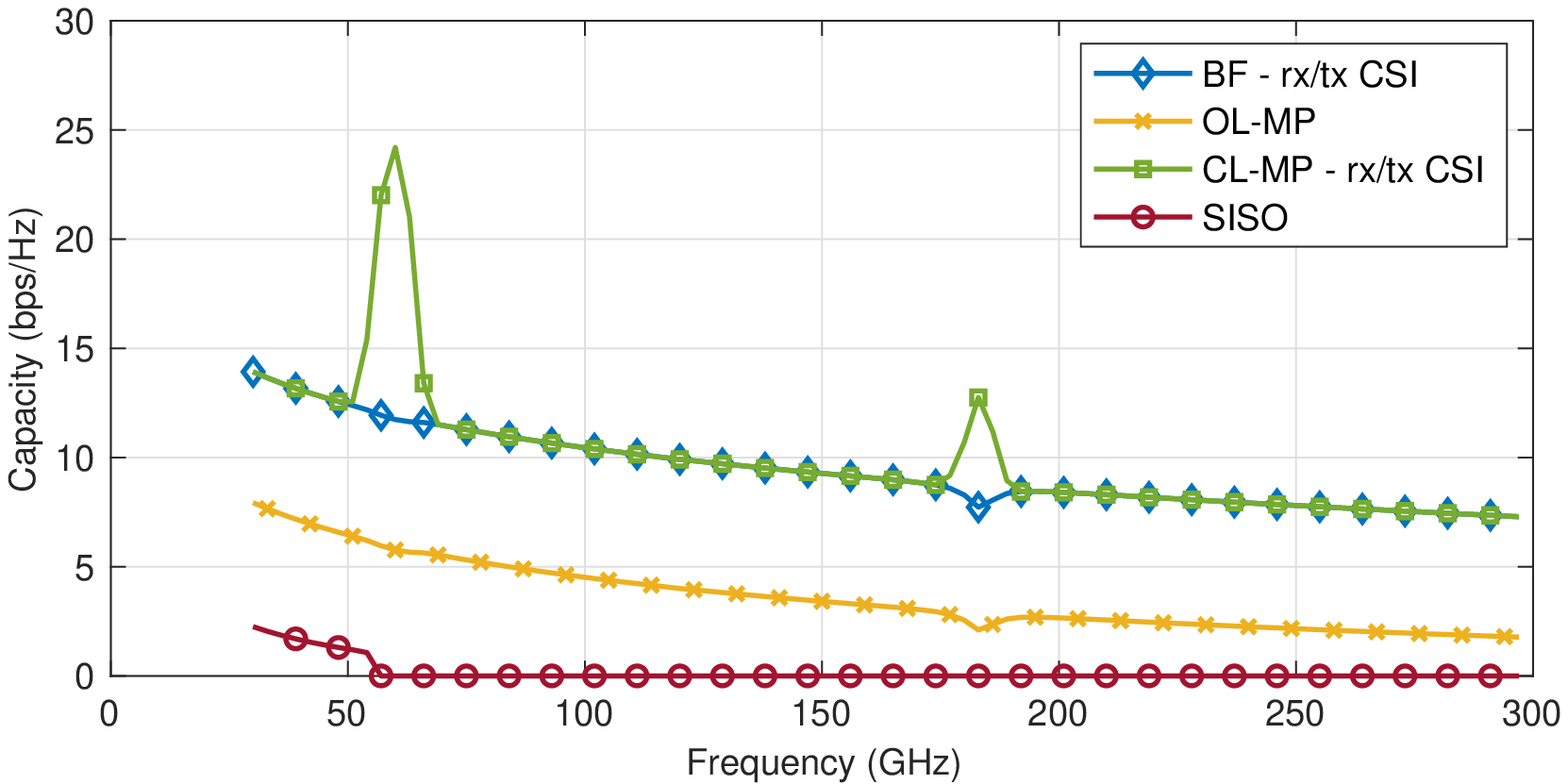}}          
        \caption{64x64 MIMO channel performance over mmWave band for different precoding techniques. The transmit power is 150\,mW. }    
        \label{fig:mmWaveResaults2}
    \end{center}
\end{figure}

Numerically speaking,
we can calculate the capacity of a SISO channel,
which turns out to be 5\,bps/Hz.
According to the existing MIMO theory,
for a full-rank 64x64 MIMO channel with enough spatial multiplexing gain,
the theoretical capacity will be expected to be $64 \times 5 \simeq 320$\,bps/Hz.
As discussed before,
the LoS MIMO channel suffers from poor spatial multiplexing gain and can achieve the maximum capacity only with some specific geometry configuration~\cite{sarris2005maximum},
which is not feasible for mobile communications.
Thus,
it can be seen that MIMO capacity, when re-radiation is assumed as noise, is close to that of a SISO channel.
However,
if the molecular re-radiation is taken into account as scattering,
it can equivalently create a rich scattering environment,
and in turn, increase the spatial multiplexing gain.

In summary,
the mmWave MIMO system can take advantage of the molecular re-radiation to generate more capacity if the re-radiation can be exploited as NLoS signal,
which prevails the absorption attenuation.

\subsection{Impact of MIMO Techniques in mmWave}

In \figurename~\ref{fig:mmWaveResaults2}, we compare the beamforming and multiplexing performance over the mmWave spectrum, when
re-radiation is considered as scattering. Four different schemes are involved: optimal beamforming (BF), Closed-Loop multiplexing with water-filling power allocation (CL-MP), and Open-Loop multiplexing with uniform power allocation (OL-MP).

It is observed that the Closed-Loop multiplexing technique results in a notably higher capacity than beamforming at high absorption frequency windows. This is because re-radiation acts as scattering and increases the multiplexing gain.

\begin{figure}[t]
	\centering
	\includegraphics[width=0.5\textwidth ,clip=true, trim=0 0 0 18]{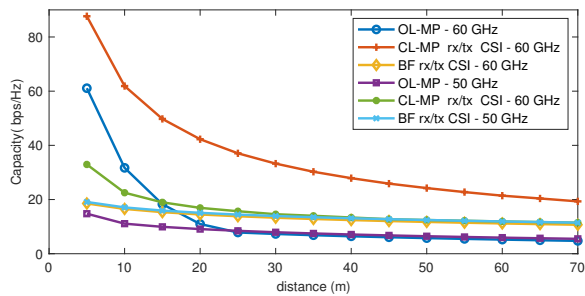}
	\caption{MIMO capacity using different techniques over communication distance. Total transmit power is 150 mW The re-radiation is assumed as scattering.}
	\label{fig:capDistmmWave}
\end{figure}

\begin{figure}[h]
	\begin{center}
		\subfloat[absorption coefficient, temperature= $273~K$, pressure= $1~atm$.]{\label{fig:kflogTHz}
			\includegraphics[width=0.5\textwidth ,clip=true, trim=0 0 0 25]{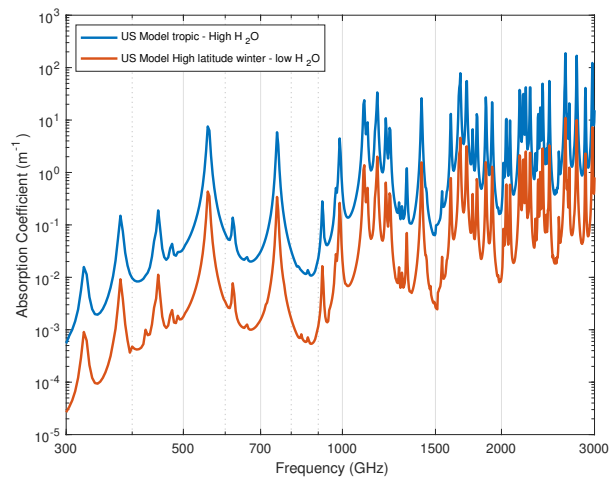}}\vspace{-0.2em}
		\subfloat[Signal Attenuation in tropic atmosphere.]{\label{fig:att}
			\includegraphics[width=0.5\textwidth ,clip=true, trim=0 0 0 0]{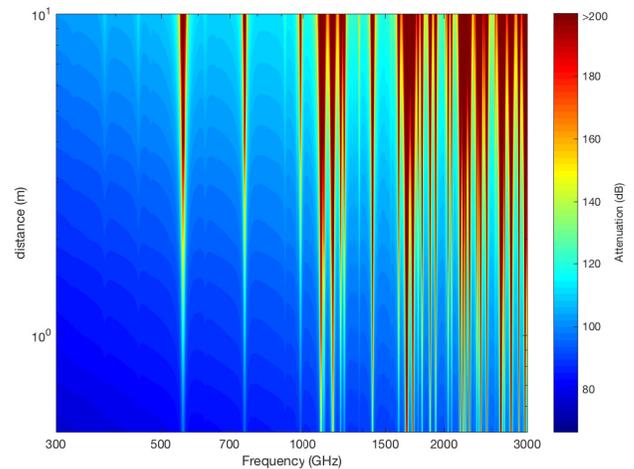}}
		\caption{Molecular absorption in terahertz band.}    
		\label{fig:abs_thz}
	\end{center}
\end{figure}

Besides, we can see that the performance of Open-Loop multiplexing is also affected by the absorption. Hence, we can see that at 60\,GHz and 180\,GHz with a distance of 5\,m, OL-MP outperforms the beamforming scheme. However, this upper hand is not observed for the larger distance when uniform power allocation is not efficient because of the low SNR.
Furthermore, a slight dip can be observed at 180\,GHz in \figurename~\ref{fig:cap35m150mW}, which shows that on the other hand, the molecular absorption weakens the LoS signal.

In \figurename~\ref{fig:capDistmmWave}, we present the performance impact of transmission distance at 50\,GHz and 60\,GHz, i.e., the low absorption spectrum vs. the high absorption spectrum.
Eq.~\eqref{eq:k-factor} shows that the longer distance results in more scattering due to more molecules in the channel, but a large path loss mitigates the advantage of scattering. Therefore, at 60\,GHz, it can be seen that although OL-MP outperforms beamforming in a short-distance channel, it loses its superiority when the distance is large. It can be explained as a consequence of allocating uniform power to all equivalent channels which mostly cannot meet the SNR threshold at a larger distance. Also, it can be observed that there is not such a quick decrease at 50\,GHz since there is not much advantage of absorption in this frequency.

\subsection{MIMO Performance in the terahertz band}
\label{sec:MIMOTHz}

\begin{figure*}[]
	\begin{center}         
		\subfloat[distance\,=\,1\,m, transmit power\,=\,1\,mW]{\label{fig:cam1m1mW}
			\includegraphics[width=0.45\textwidth]{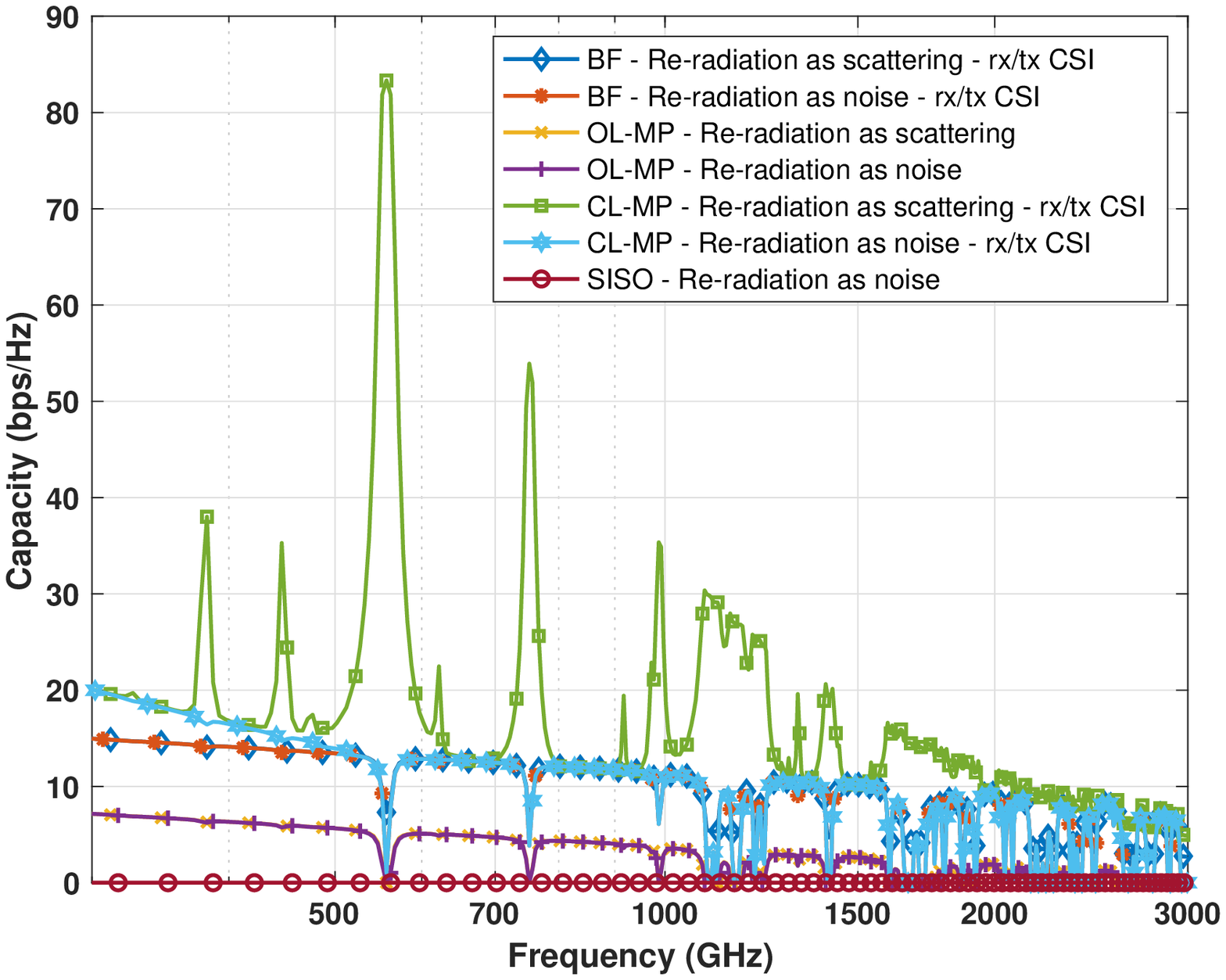}}   
		\subfloat[distance\,=\,1\,m, transmit power\,=\,10\,mW]{\label{fig:cam1m10mW}
			\includegraphics[width=0.45\textwidth ,clip=true, trim=0 0 0 0]{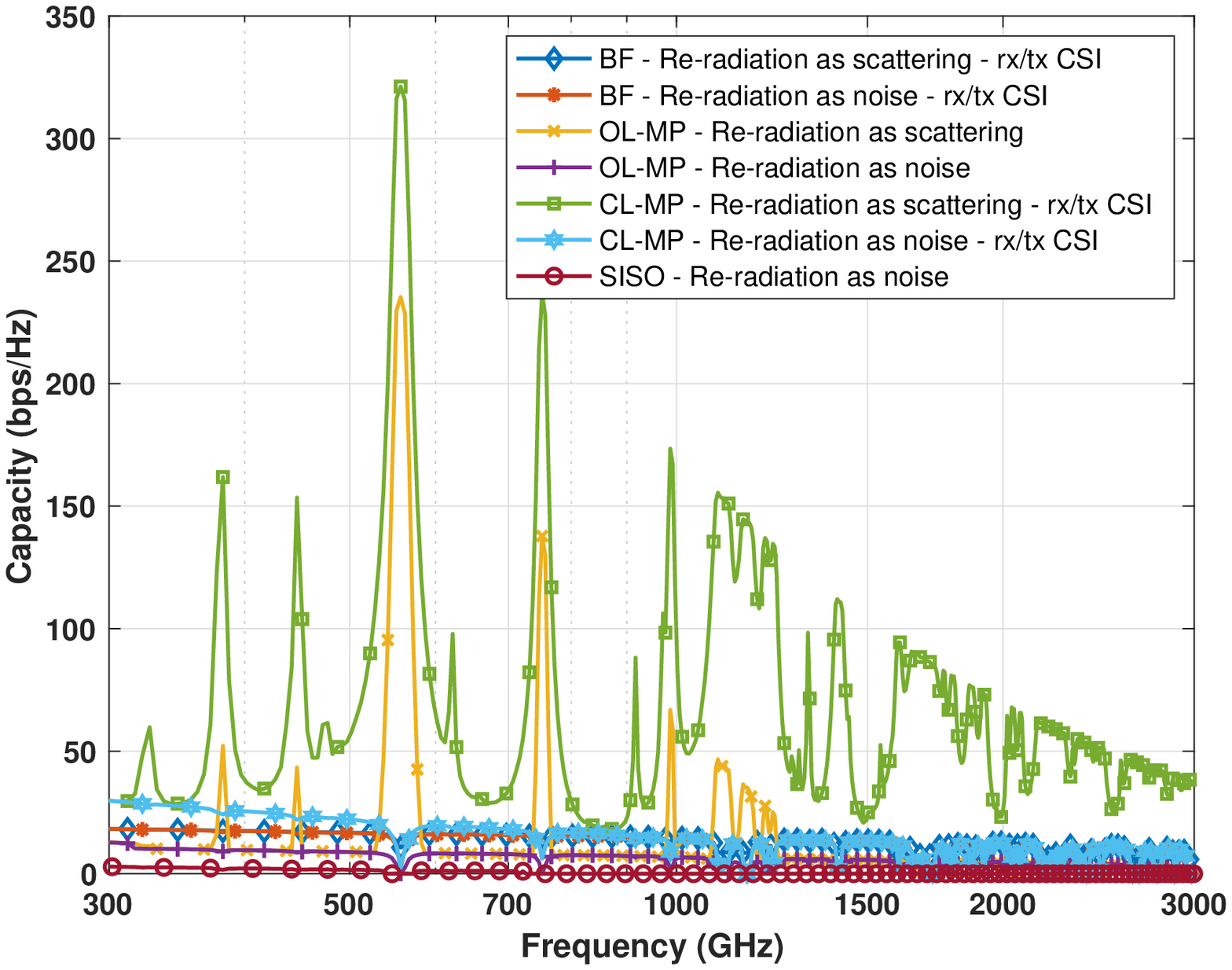}}\vspace{-1em}  
		\subfloat[distance\,=\,10\,m, transmit power\,=\,1\,mW]{\label{fig:cam10m1mW}
			\includegraphics[width=0.45\textwidth ,clip=true, trim=0 0 0 0]{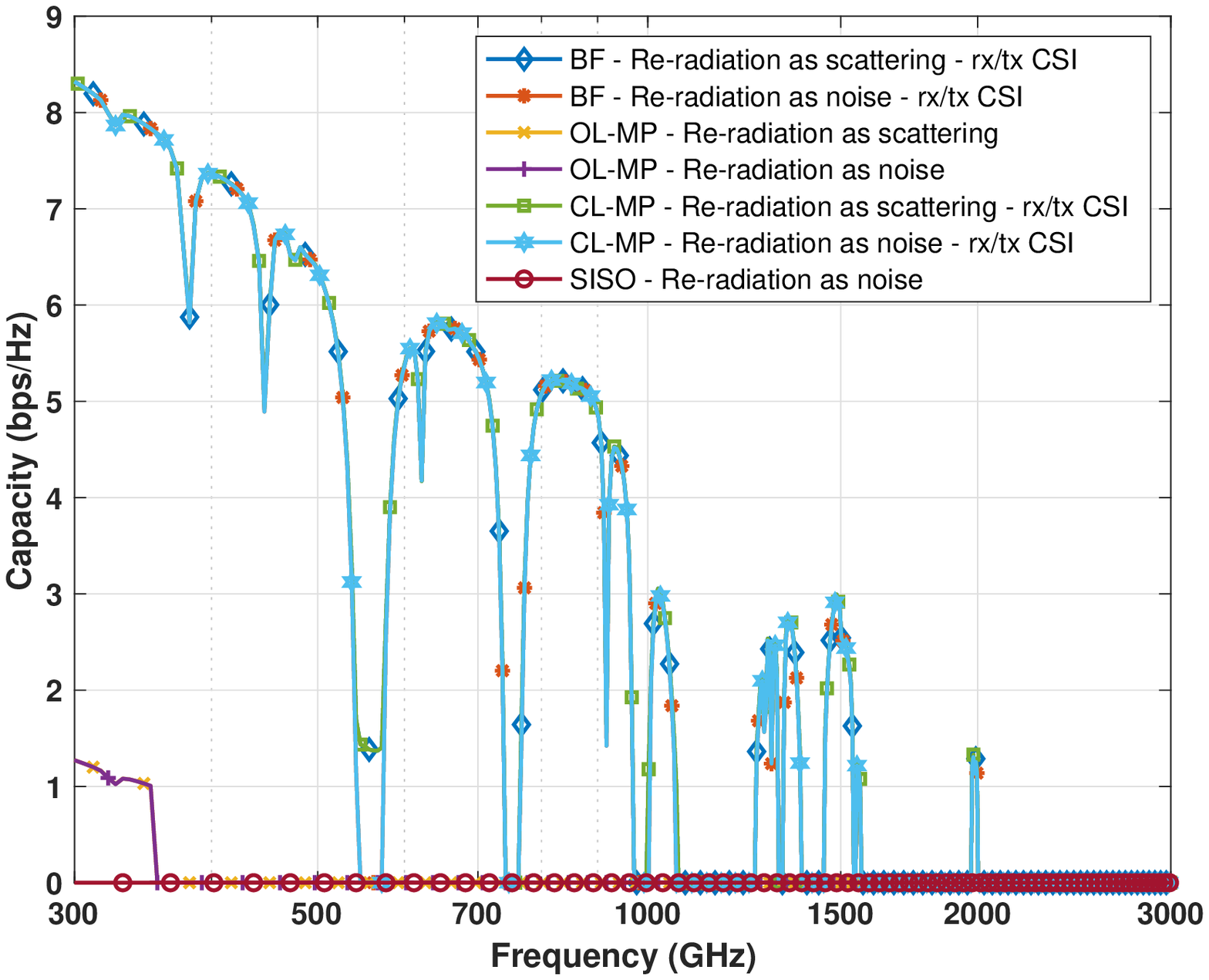}}
		\subfloat[distance\,=\,10\,m, transmit power\,=\,10\,mW]{\label{fig:cam10m10mW}
			\includegraphics[width=0.45\textwidth]{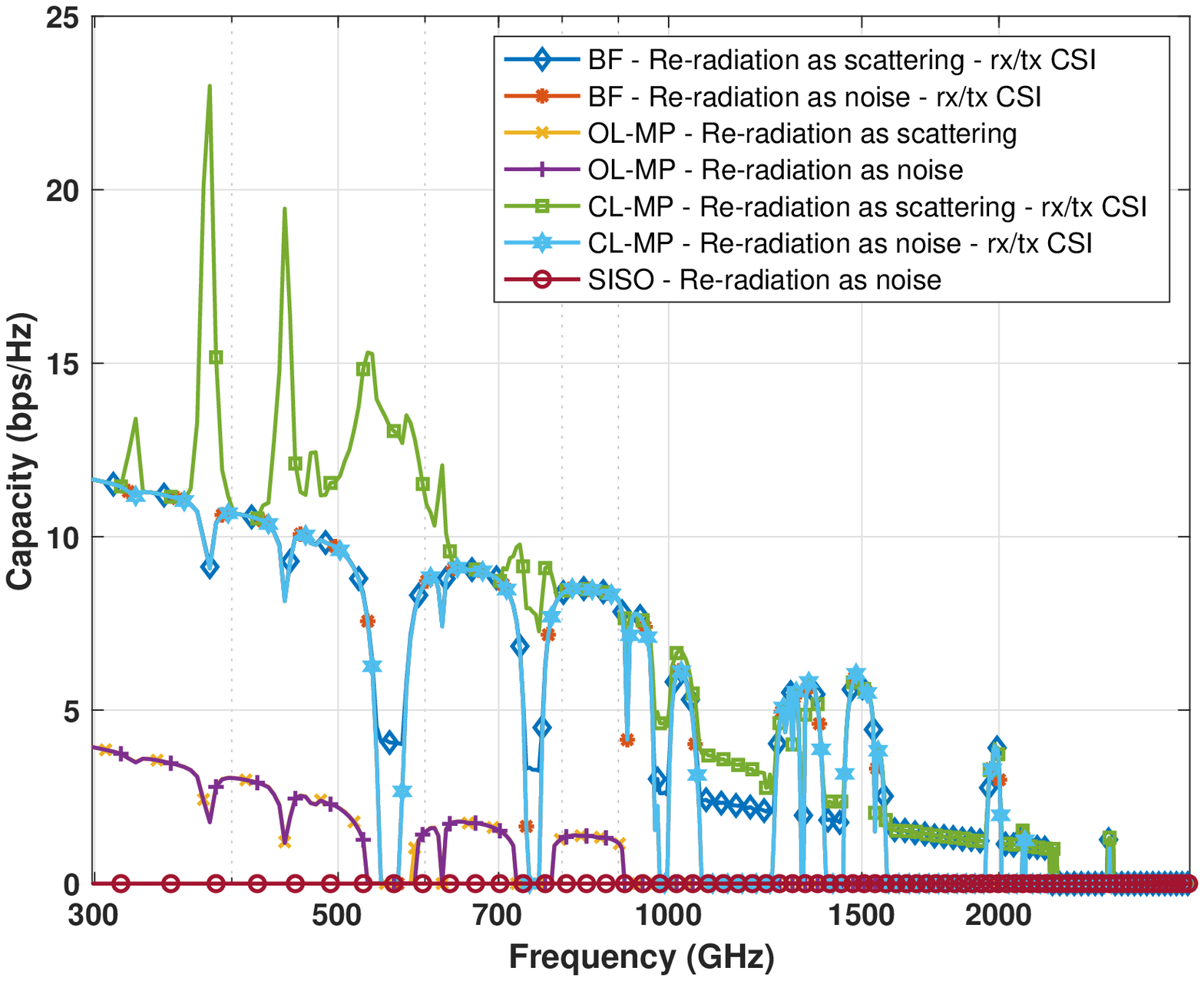}}
		\vspace{-0.2\baselineskip}           
        \caption{225x225 MIMO channel performance over terahertz band for various precoding techniques.}    
        \label{fig:cap1}
    \end{center}
\vspace{-0.5cm}
\end{figure*}

In this section, we extend our investigation to the terahertz band. The channel is simulated with different transmit power and distances. In the terahertz band, due to very high propagation loss, the applications are limited to short-range communications. Our assumption on the transmit power is based on current technology \cite{Akyildiz201416} and a previous work on terahertz massive MIMO \cite{AKYILDIZ2016massive}. Moreover, several channel distances are selected to cover various nominated terahertz applications. For example, terahertz nano-sensors are supposed to communicate in a very short distance in the order of 0.1-10 cm or less, while terahertz communications are also nominated to provide terabit per second ultra-high-speed video communication link at around 1\,m distance for home entrainment devices like TV or virtual reality (VR)~\cite{chaccour2019reliability}. Terahertz application is also extended to wireless personal or local networks where the channel distance is up to a few meters. Since the application of MIMO in nano-sensors communication is rare, we limit the simulation to distances 1 and 10\,m.

In the terahertz band, the dominant absorption source in the air is water molecules. We can see in \figurename~\ref{fig:kflogTHz} that all absorption peaks in the tropic atmosphere are higher than the winter atmosphere. The main difference is the mole ratio of vapor molecules in the air. The channel attenuation including molecular attenuation in~\eqref{eq:Atten_abs} and FSPL attenuation in~\eqref{eq:Atten_Spread} is illustrated in \figurename~\ref{fig:att}. While the FSPL attenuation is increasing linearly (in\,dB) over distance and frequency, the molecular attenuation is also increasing with distance but it is frequency selective. 

\balance

\figurename~\ref{fig:cam1m1mW} and~\ref{fig:cam1m10mW} illustrate the capacity of the MIMO techniques with a 1\,m distance. The transmit power is increased from 1\,mW in \figurename~\ref{fig:cam1m1mW} to 10\,mW in \figurename~\ref{fig:cam1m10mW}. Similar to mmWave when molecular re-radiation is considered as noise, molecular attenuation and noise lead to a lower capacity in the high absorption frequency windows for all precoding techniques. It follows the current understanding of the terahertz channel and we can see regardless of the MIMO technique, the capacity drops sharply in high absorption windows. On the other hand, results change drastically when molecular re-radiation is assumed to be scattering. In this case, the LoS channel converts to a Rician or Rayleigh channel depending on the channel absorption coefficient. As expected, multiplexing results in a much better performance than beamforming since it can employ spatial multiplexing. For example, it can be observed in \figurename~\ref{fig:cam1m1mW} that CL-MP outperforms the beamforming thanks to the tremendous multiplexing gain provided by the rich scattering environment due to molecule re-radiation.
In more detail, a significant capacity improvement can be observed at very high absorption frequencies such as 540-560\,GHz. Communication over such frequency windows is considered infeasible for terahertz communications in existing studies. While at 500\,GHz the beamforming and multiplexing capacity are 15 and 17\,bps/Hz respectively, those are about 7 and 84\,bps/Hz at 550\,GHz. Thus, the absorption and re-radiation transforms the LoS dominant channel to a Rayleigh channel and reduces beamforming performance. The reason can be found in Section~\ref{sec:analysis}, where we have discussed that re-radiation decreases the K-factor and creates a rich scattering channel. 

However, it is also observed in \figurename~\ref{fig:cam1m1mW} that the OL-MP results in poor performance in comparison to CL-MP and beamforming. The power allocation scheme downgrades to uniform power allocation where an identity matrix is used for precoding due to the lack of CSI at the transmitter. Thus, the capacity is equal to that of the re-radiation-as-noise case and drops to zero in high absorption windows. When the power is distributed uniformly, the equivalent SNR, \large ($\frac{P\lambda_i^2}{m\sigma^2}$)\normalsize, of most parallel channels is less than 0~dB. Practically, the receiver cannot detect the transmitted symbol when the signal is weaker than noise. These results are matched with the existing literature that the open-loop multiplexing performance drops dramatically in the low SNR~\cite{gesbert2003theory}.

\figurename~\ref{fig:cam1m10mW} illustrates a high SNR scenario. It shows the CL-MP and OL-MP result in a close capacity, significantly higher than beamforming at very high absorption frequencies such as 550\,GHz. This is because the uniform power allocation is close to optimum in a very rich scattering channel~\cite{Jayaweera2002Rician}. In other words, the water-filling scheme results in a uniform power allocation.

In summary, considering the same implementation challenges of beamforming and CL-MP, OL-MP might still be a preferable choice for frequency up to 1\,THz.
But when there is not enough signal strength at the receiver, the transmitter should have CSI to steer the beam toward the receiver. Furthermore, the advantage of CL-MP in comparison with beamforming is that the former can take advantage of both beam shaping and multiplexing over parallel paths. However, CSI overhead can be very large for massive MIMO. For example, the channel transfer matrix has 50625 elements with a 225x225 MIMO system where each element is a complex value and required at least 16 bits of data. Nevertheless, the channel coherent time in presence of molecular re-radiation has not been investigated. The updating interval of CSI feedback should be significantly shorter than the channel coherent time to avoid using outdated CSI~\cite{Banelli2014OFDM}. Thus, having the non-expired full CSI at the transmitter can decrease the effective spectral efficiency of the reverse link~\cite{Sibille2011LTEbook}.

\section{Conclusion}
\label{sec:con}

In this paper, we have presented a new perspective on molecular absorption and re-radiation over the mmWave and terahertz band. We reviewed two alternative assumptions on the effects of molecular re-radiation, noise versus scattering. While re-radiation has been mostly considered as noise in the literature, we have also considered the recent idea that the re-radiation is correlated to the main signal and thus can be modeled as scattering. Hence, we combined both theories and characterized a multi-path channel which assumes the molecular re-radiation as scattering. Our simulation results showed that the re-radiation can greatly improve MIMO performance in the mmWave/terahertz spectrum when it is treated as scattering. We have shown that the frequency windows, which were previously considered as very high attenuation sub-bands, are actually more efficient in MIMO as they can take advantage of molecular re-radiation. Our results have further confirmed that multiplexing can be a viable alternative to beamforming even in an LoS channel, which might fundamentally changes the conclusion drawn from the traditional MIMO communications theory.

\section{Future Work}
\label{sec:fut}

The theoretical discovery in our research can potentially change our understanding of wireless communications over the mmWave/terahertz band. Future work would focus on the experimental measurements, which requires a set of well-designed measurement scenarios to detect the re-radiation and show its correlation with LoS signal. To investigate the re-radiation,
a high and a low absorption scenarios can be experimented. This can be realized through manipulating the density of channel molecules, e.g. oxygen at 60\,GHz, or measuring in slightly different frequencies where the absorption is significantly different, e.g. 57, 60 and 63\,GHz. 
Furthermore, the re-radiated signal characteristics should also be measured and characterized. Finally, the performance of a MIMO system should be measured with various precoding schemes.

The experimental future work, however, must overcome several challenges: 1) the high-frequency equipment used in the experiments may not be sensitive enough to detect the re-radiated energy, 2) the detection of phase is difficult for high-frequency signals, especially when other factors such as wall and roof reflection also affect the signal propagation, and 3) it may be difficult to control the oxygen and vapor of the communication medium, especially considering the very limited absorption changes over environment parameters.

\bibliographystyle{IEEEtran}
\balance
\bibliography{main}
\begin{IEEEbiography}
    [{\includegraphics[width=1in,height=1.25in,clip,keepaspectratio]{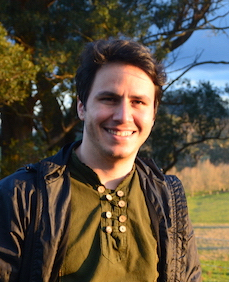}}]{Sayed Amir Hoseini} is a postdoctoral research associate at the University of New South Wales at ADFA-Canberra. He received the BSc degree in Electronic Engineering from the Isfahan University of Technology and the MSc degree in Electronic and Communication Engineering from the Amirkabir University of Technology (Tehran Polytechnic), in Iran in 2008 and 2011, respectively. He completed PhD in Computer Science and Engineering at the University of New South Wales (UNSW Sydney) in 2017. He worked at CSIRO\textbar DATA61 and Central Queensland University as a postdoctoral researcher. Since mid-2020, he has joined School of Engineering and Information Technology at UNSW-Canberra. His research interests include Wireless Communications, Physical Layer Security, and UAV Communication.
\end{IEEEbiography}
\vskip 0pt plus -1fil
\begin{IEEEbiography}
    [{\includegraphics[width=1in,height=1.25in,clip,keepaspectratio]{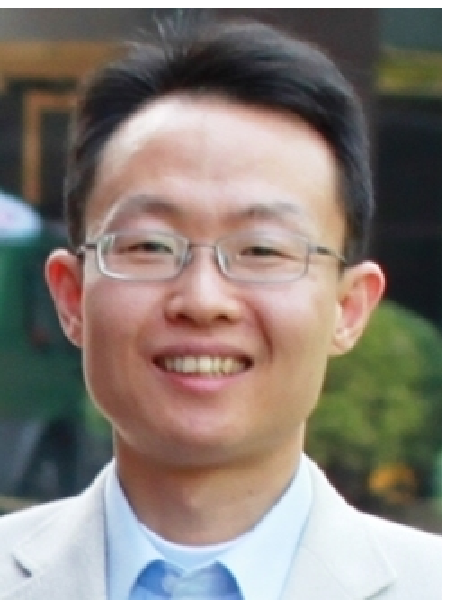}}]{Ming Ding} (M’12-SM’17) received the B.S. and M.S. degrees (with first-class Hons.) in electronics engineering from Shanghai Jiao Tong University (SJTU), Shanghai, China, and the Doctor of Philosophy (Ph.D.) degree in signal and information processing from SJTU, in 2004, 2007, and 2011, respectively. From April 2007 to September 2014, he worked at Sharp Laboratories of China in Shanghai, China as a Researcher/Senior Researcher/Principal Researcher. He also served as the Algorithm Design Director and Programming Director for a system-level simulator of future telecommunication networks in Sharp Laboratories of China for more than 7 years. Currently, he is a senior research scientist at Data61, CSIRO, in Sydney, NSW, Australia. His research interests include information technology, data privacy and security, machine learning and AI, etc. He has authored over 100 papers in IEEE journals and conferences, all in recognized venues, and around 20 3GPP standardization contributions, as well as a Springer book “Multi-point Cooperative Communication Systems: Theory and Applications”. Also, he holds 21 US patents and co-invented another 100+ patents on 4G/5G technologies in CN, JP, KR, EU, etc. Currently, he is an editor of IEEE Transactions on Wireless Communications and IEEE Wireless Communications Letters. Besides, he has served as Guest Editor/Co-Chair/Co-Tutor/TPC member for many IEEE top-tier journals/conferences and received several awards for his research work and professional services.
\end{IEEEbiography}
\vskip 0pt plus -1fil
\begin{IEEEbiography}
    [{\includegraphics[width=1in,height=1.25in,clip,keepaspectratio]{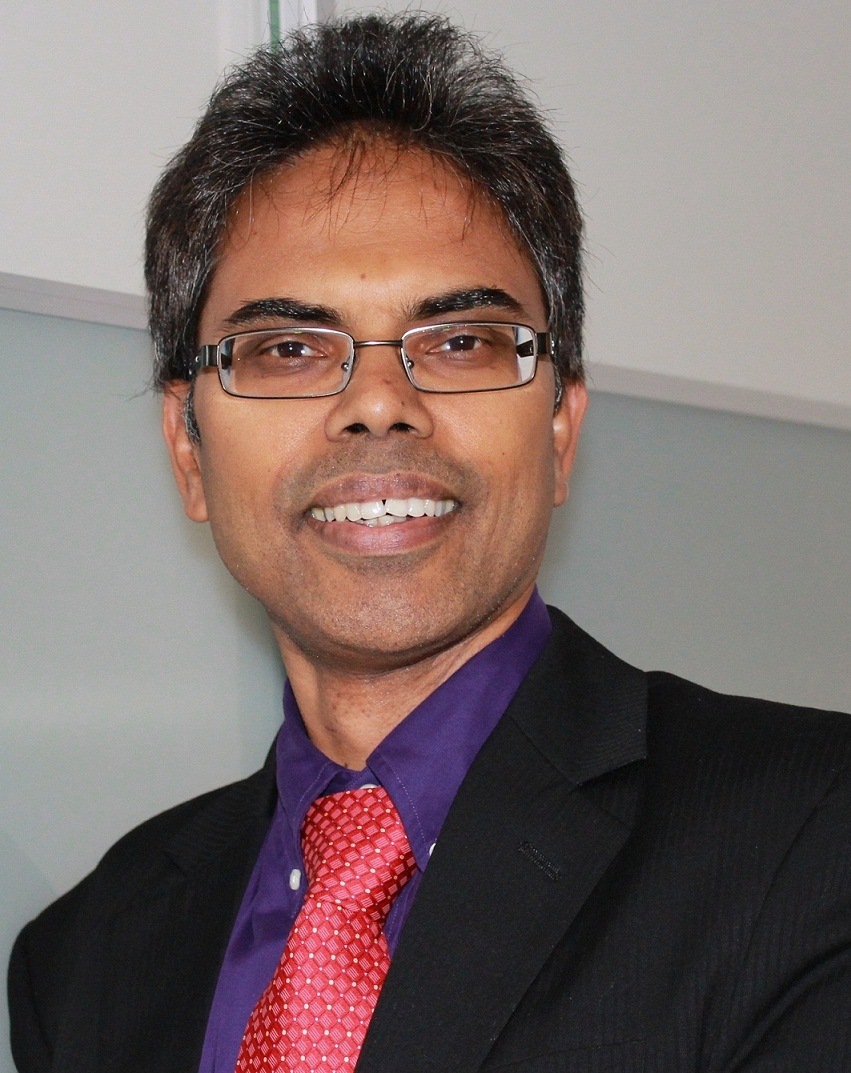}}]{Mahbub Hassan} (SM'00) is a Full Professor in the School of Computer Science and Engineering, University of New South Wales, Sydney, Australia. He has PhD from Monash University, Australia, and MSc from University of Victoria, Canada, both in Computer Science. He served as IEEE Distinguished Lecturer and held visiting appointments at universities in USA, France, Japan, and Taiwan. He has coauthored three books, over 200 scientific articles, and a US patent. He served as editor or guest editor for many journals including IEEE Communications Magazine, IEEE Network, and IEEE Transactions on Multimedia. His current research interests include Mobile Computing and Sensing, Nanoscale Communication, and Wireless Communication Networks. More information is available from {http://www.cse.unsw.edu.au/\texttildelow mahbub/}.
\end{IEEEbiography}
\nobalance
\begin{IEEEbiography}[{\includegraphics[width=1in,height=1.25in,clip,keepaspectratio]{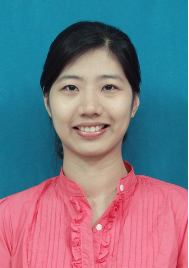}}]{Youjia Chen} (M'18) received the B.S. and M.S degrees in communication engineering from Nanjing University, Nanjing, China, and Ph.D. degree in wireless engineering from the University of Sydney, Australia, in 2005, 2008 and 2017, respectively. From 2008-2009, she worked in Alcatel-Lucent Shanghai Bell. Then she worked at the College of Photonic and Electrical Engineering, Fujian Normal University, China, from Aug. 2009. In 2018, she joined the College of Physics and Information Engineering, Fuzhou University, China. Her current research interests include ultra-dense networks, wireless caching and deep learning in wireless networks.
\end{IEEEbiography}
\end{document}